\documentclass[journal,onecolumn]{IEEEtran} 

\usepackage[T1]{fontenc} 
\usepackage{enumerate}
\usepackage{colortbl}
\usepackage{amsmath}
\usepackage{amsfonts}
\usepackage{mathrsfs}
\usepackage{booktabs}
\usepackage{graphicx}
\usepackage[subfigure]{graphfig}
\usepackage[latin1]{inputenc}
\usepackage{epstopdf}
\usepackage{amsmath,bm}
\usepackage[numbers,sort&compress]{natbib}

\usepackage{natbib}

\begin{document}

\title{Process monitoring based on orthogonal locality preserving projection with maximum likelihood estimation}

\author{~Jingxin Zhang,  Maoyin Chen, Hao~Chen, Xia Hong, and Donghua Zhou 
\thanks{ingxin Zhang and Maoyin Chen are with Department of Automation, TNList, Tsinghua University, Beijing, China.(e-mail: zjx18@mails.tsinghua.edu.cn)}
\thanks{Xia Hong is with the Department of Computer Science, School of Mathematical, Physical and Computational Sciences, University of Reading, RG6 6AY, U.K.}
\thanks{Donghua Zhou is with College of Electrical Engineering and Automation, Shandong University of Science and Technology, Qingdao, China and Department of Automation, TNList, Tsinghua University, Beijing, China {zdh@mail.tsinghua.edu.cn}}
\thanks{This paper has been published in Industrial \& Engineering Chemistry Research}}

\maketitle

\begin{abstract}
By integrating two powerful methods of density reduction and intrinsic dimensionality estimation, a new data-driven method, referred to as OLPP-MLE (orthogonal locality preserving projection-maximum likelihood estimation), is introduced for process monitoring.  OLPP is utilized for dimensionality reduction, which provides better locality preserving power than locality preserving projection. Then, the MLE is adopted to estimate intrinsic dimensionality of OLPP.  Within the proposed OLPP-MLE, two new static measures for fault detection $T_{\scriptscriptstyle {OLPP}}^2$ and ${\rm SPE}_{\scriptscriptstyle {OLPP}}$ are defined.  In order to reduce algorithm complexity and ignore data distribution, kernel density estimation is employed to compute thresholds for fault diagnosis. The effectiveness of the proposed method is demonstrated by three case studies.
\end{abstract}

\section{Introduction}\label{introduction}

In recent decades, process monitoring becomes an increasingly significant and valuable research topic due to the high requirements of safety and reliability in industrial applications~\cite{shang2018recursive,Musulin2014Spectral,Shi2018Fault}.  Specularly, data-driven process monitoring has attracted worldwide attention and has acquired remarkable accomplishments~\cite{LWPR,shang2018isolating,zhou2018distributed,yang2019using}. The key contribution of data-driven techniques is  to take advantage of sensing variables to detect faults, which makes them applicable in realistic industrial systems~\cite{jiang2016bayesian,Shang2017Recursivea,Molanes2018Deep}.In modern society, industrial systems such as power plants and high-speed rail become more complex, and produce massive data even in just one hour.   Hence, how to extract valuable information from available data becomes the most critical issue at present.

There are various  methods for  data-driven process monitoring now. Classical schemes include principal component analysis (PCA)~\cite{comparison}, partial least squares (PLS)~\cite{zhou2010total,TE2}, independent component analysis (ICA)~\cite{constrained}, canonical variate analysis~\cite{kde}, etc.  Owing to simplicity and effectiveness in dealing with large data, PCA is recognized as a popular dimensionality reduction technique for linear systems, which has been widely applied to feature extraction as well as process monitoring areas~\cite{li2014new,Yu2016An}. For instance, recursive total principle component regression was proposed for vehicular cyber-physical systems, which is able to detect small faults and suitable for on-line implementation~\cite{jiang2018recursive}. There are also numerous fault detection methods based on PCA and PLS~\cite{TE2,yin2016an}, which have been integrated on the Matlab toolbox \cite{Jiang2016Comparison}. However, PCA aims to discover the global geometric topological structure of the Euclidean space, without considering the underlying local manifold structure.

Locality preserving projection (LPP) is regarded as an effective way to replace PCA, where local neighborhood framework of data could be optimally preserved~\cite{lpp,olpp}. Based on LPP, orthogonal locality preserving projection (OLPP) was proposed to reconstruct data conveniently, where mutually orthogonal basis functions are calculated. Besides, OLPP can realize the function of PCA when the parameter is set appropriately and eigenvectors corresponding to largest eigenvalues are retained\cite{he2005face}, which indicates that OLPP can not preserve global and local geometric structure simultaneously.

Note that OLPP shares several data representation characteristics with nonlinear techniques, for instance, Laplacian Eigenmaps (LE)~\cite{LE}, Isomap~\cite{Isomap} and locally linear embedding~\cite{LLE,Mcclure2013Nonlinear}. These local nonlinear manifold approaches are non-parametric without parametric hypothesis and are casted into the eigen-problem instead of iteration, which makes them considerably less complicated~\cite{tu2012Laplacian,zhang2013isomap}. However, ``out of sample'' issue severely constraints the applications of process monitoring.  OLPP is exactly the linear extension of LE algorithm and can efficiently deal with this issue.

For the OLPP approach, the intrinsic dimensionality (ID) is the most critical parameter~\cite{hasanlou2012comparative}.
If the ID is too small, significant data characteristics may be ``collapsed'' onto the same dimension. However, if the ID is too large, the projections become noisy and may be unstable. Therefore, the estimation of the optimal ID is the prime task that should be considered.
In this paper, as a local estimator of ID,  maximum likelihood estimation (MLE) is adopted to calculate the ID with little artificial interference, which has the superiorities of easy-implementation, stability, high reliability~\cite{mle,kegl2002intrinsic,f2003data}. Besides, it is also robust to noisy data and less computationally complicated on high dimensional data~\cite{Carter2010On}.

Since OLPP is a preferable choice for dimensionality reduction, an improved data-driven process monitoring approach is proposed based on OLPP, referred as to OLPP-MLE, where MLE is embedded in OLPP framework. Unfortunately, although the ID estimator is provided, OLPP still encounters the singular issue frequently, especially for data with zero mean. Within the proposed approach, the singular problem is efficiently settled with three optional solutions, which is the meaningful improvement compared with traditional OLPP. Besides, an alternative manner is provided to calculate eigenvector with little computational cost.

The virtues of the proposed OLPP-MLE are summarized as follows:

\begin{enumerate}[a)]
  \item The proposed approach provides more locality preserving power and discriminating power than most typical dimensionality reduction approaches;

  \item Because OLPP and MLE are based on local geometric characteristics of data, the proposed approach is considerably less computationally complicated in contrast with global approaches;

  \item  For the potential singular problem within the standard OLPP, three alternative solutions are summarized, which provides more choices for researchers to select;

  \item  MLE provides a stable estimation of the ID, which is insensitive to parameter tuning;

  \item The proposed method has no requirement of data distribution, which is beneficial to expand its applications.
\end{enumerate}

The remaining parts of this paper are organized below. Section \ref{olpp-basic} summarizes the preliminary of OLPP and MLE on intrinsic dimensionality briefly. Major procedure of OLPP-MLE approach is summarized for data-driven process monitoring in Section \ref{olpp-mle}.
The solutions of singular problem, an alternative approach to calculate eigenvectors and computational complexity analysis are also discussed thereafter. Then, a numerical case study and the continuous stirred tank reactor (CSTR) are adopted to illustrate the effectiveness of the proposed approach in Section \ref{numerical-case}. Section \ref{section5} utilizes  Tennessee Eastman (TE) process to verify the stability of the proposed approach and  to compare with several typical data-driven approaches. Conclusion is presented in Section \ref{conclusion}.

\section{Preliminaries}\label{olpp-basic}
In this section, we introduce two independent important works on MLE of ID~\cite{mle} and the algorithm of OLPP~\cite{olpp}, which are both based on the local geometric properties of data. These works form the building stones of our proposed data-driven process monitoring algorithm.

\subsection{Maximum likelihood estimation on intrinsic dimensionality}\label{determination-intrinsic}
The MLE of ID is derived by ~\citeauthor{mle} \cite{{mle}}. Assume a data set ${\{\boldsymbol x_i\}_{i=1}^N \subset {\mathcal R^m}}$, representing an embedding of a lower-dimensional sample $ \boldsymbol{x_i} =g(\boldsymbol {y_i})$, where $g(.)$ is a continuous and smooth mapping, and $\boldsymbol y_i$ are sampled from an unknown density $p$ on ${\mathcal R^l}$. For the purpose of ID estimation, it is initially assumed that $p\approx const$ in a small hyper sphere $H_{\boldsymbol{x}}(t)$ around a data point $\boldsymbol x$.  The data points inside the hyper sphere are modelled as a  Poisson process. Let the counts of the observations with distance $t$ from $\boldsymbol x$ be denoted as
\begin{equation}
C(t, \boldsymbol {x})= \sum_{i=1}^{N}1\{ \boldsymbol{ x}_i \in H_{\boldsymbol{x}}(t)   \}
\end{equation}
where $0 \le t \le R$, $H_{\boldsymbol{x}}(t)$  is a small hyper sphere around a data point $\{\boldsymbol x\}$ with radius $t$. For fixed $N$, $C(t,\boldsymbol x)$ is   approximated as a Poisson process, and the rate of the process $\lambda \left( t \right)$ of the $C(t,\boldsymbol x)$ is  given by
  \begin{equation}\label{dim1}
    \lambda \left( t \right) =p(\boldsymbol{x})V(l)   l \cdot t^{l-1},
  \end{equation}
 where $V(l)=\pi^{l/2}[\Gamma(l/2+1)]^{-1}$ denotes the volume of the unit sphere in ${\mathcal R^l}$.
 It can be shown that the log-likelihood of the observed $C(t,\boldsymbol x)$ can be expressed as
 \begin{equation}\label{ctx}
 \int_0^R \log \lambda (t) d C(t) -\int_0^R \lambda(t) dt,
 \end{equation}
 The maximization of (\ref{ctx}) results in a unique solution $l$. In practice, $l$ can be obtained based on a data point $\boldsymbol x_i$ given $k$ nearest neighbors, which is calculated by~\citep{mle}
  \begin{equation}\label{mle-intrinsic}
{\mathop l\limits^ \wedge}  _k\left( {{\boldsymbol x_i}} \right) = {\left( {\frac{1}{{k - 1}}\sum\limits_{j = 1}^{k - 1} {\log \frac{{{F_k}\left( {{\boldsymbol x_i}} \right)}}{{{F_j}\left( {{\boldsymbol x_i}} \right)}}} } \right)^{ - 1}},
  \end{equation}
 where ${{F_k}\left( {{\boldsymbol x_i}} \right)}$ is the smallest radius of the hypersphere with center $\boldsymbol x_i$, which must contain $k$ neighboring data points.

It is obvious that $k$ affects the estimate severely. In general, the estimator is expected to be small enough and contain as many points as possible. In our approach, just average over a range of small to moderate values $k=k_1,\ldots,k_2$ to obtain the optimal estimate
\begin{equation}\label{estimatel}
 {\widehat l_k} = \frac{1}{N}\sum\limits_{i = 1}^N {{{\widehat l}_k}\left( {{\boldsymbol{x}_i}} \right)} ,\quad \widehat l = \frac{1}{{{k_2} - {k_1} + 1}}\sum\limits_{{k_1}}^{{k_2}} {{{\widehat l}_k}}.
\end{equation}

The perfect range $k_1,\ldots,k_2$ is different for every combination of $l$ and $N$, but the estimation of dimensionality is considerably stable, which has already been demonstrated~\cite{mle}. Given $l$ and $N$, $l$ is not sensitive to $k_1$ and $k_2$, which is discussed in Section \ref{discussionofl}. For simplicity and reproducibility, $k_1$ and $k_2$ are fixed in this paper.

\subsection{Orthogonal locality preserving projection}
OLPP is a common dimensionality reduction approach, and can preserve the local geometric characteristics of the manifold~\cite{olpp}.
Given a data set $\boldsymbol{X}=  [{\boldsymbol x_1}, ...., {\boldsymbol x_N}] $ with $\boldsymbol x_i \in \mathcal R^m, i = 1,\cdots, N$. 
Let $\boldsymbol S$ be a similarity matrix defined on data points, which is computed by
\begin{equation}\label{eq1}
  {\boldsymbol S_{ij}} = {e^{ - \frac{{{{\left\| {{\boldsymbol x_i} - {\boldsymbol x_j}} \right\|}^2}}}{q}}},
\end{equation}
where $q$ is predefined by users. Define $\boldsymbol D$  as a diagonal matrix with ${{\boldsymbol D_{ii}} = \sum\limits_{j = 1}^N {{\boldsymbol S_{ji}}}} $. Then, ${\boldsymbol L=\boldsymbol D-\boldsymbol S}$ is Laplacian matrix in graph theory. The objective function of OLPP can be written as
\begin{equation}
\{\boldsymbol{a}_1, ...,\boldsymbol{a}_l \}=\min_{\boldsymbol{a}} \{f(\boldsymbol{a} )=\frac{{\boldsymbol a}^{\rm T}\boldsymbol{X}\boldsymbol{L} \boldsymbol{X}^{\rm T}   {\boldsymbol a} }{{\boldsymbol a}^{\rm T}\boldsymbol{X}\boldsymbol{D} \boldsymbol{X}^{\rm T}   {\boldsymbol a} } \}
\end{equation}
 subject to $\boldsymbol{a}_i^{\rm T}\boldsymbol{a}_j=0$, for $i\neq j$.    The OLPP algorithm is presented concretely in Appendix. The OLPP aims at finding ${\left\{ {{\boldsymbol y_i}} \right\}_{i = 1}^N \in  {\mathcal R^l},l \ll m}$, where $\boldsymbol y_i$ can \lq\lq represent'' $\boldsymbol x_i$, and $l$ is the ID of data. OLPP is applicable especially in the particular situation, where ${{\boldsymbol x_1},{\boldsymbol x_2}, \cdots ,{\boldsymbol x_N} \in {\boldsymbol M}}$ and $\boldsymbol M$ is a nonlinear manifold.

\section{OLPP-MLE for fault diagnosis}\label{olpp-mle}

In this section, two well-known statistics  $T^2$ and $\rm SPE$~\cite{datasx}, transitionally based on PCA or other variants~\cite{ding2010on,comparison} are served as indices to monitor the operating process, referred to as $T_{\scriptscriptstyle {OLPP}}^2$ and ${\rm SPE}_{\scriptscriptstyle {OLPP}}$. As aforementioned, OLPP has more locality preserving power than most typical dimensionality reduction methods, which may lead to discriminating capability for data anomaly. 

Consider that OLPP is applied based on a set of normal training samples ${\{\boldsymbol x_i\}_{i=1}^N \in  {\mathcal R^m}}$ based on a given dimension $l$, which is obtained by MLE of the ID in Section~\ref{determination-intrinsic}. 
Let ${{\boldsymbol W_{\scriptscriptstyle {OLPI}}} = \left[ {{{\boldsymbol a}}_1, \cdots ,{ {\boldsymbol a}}_l} \right]}$, with ${{\boldsymbol W_{\scriptscriptstyle {OLPI}}^{\rm T}}{\boldsymbol W_{\scriptscriptstyle {OLPI}}}= {\boldsymbol I}}$.  The resultant OLPI mapping becomes
\begin{equation}\label{y-representation}
  \boldsymbol x \to \boldsymbol y = {\boldsymbol W}_{\scriptscriptstyle {OLPI}} ^{\rm T} {\boldsymbol x},
\end{equation}
where $\boldsymbol y$ is an $l$-dimensional expression of raw $\boldsymbol x$. Denote ${\boldsymbol{Y}=  [{\boldsymbol y_1}, ...., {\boldsymbol y_N}]}$.

$T_{\scriptscriptstyle {OLPP}}^2$  monitoring statistic for principal component subspace
is calculated by
\begin{equation}\label{eq7}
 {T_{\scriptscriptstyle {OLPP}}^2} = {\boldsymbol y^{\rm T}}{\boldsymbol \Lambda ^{ - 1}}{\boldsymbol y},
\end{equation}
where the elements of ${\boldsymbol \Lambda  = diag({\gamma_1}, \ldots ,{\gamma_l})}$ are the eigenvalues of the covariance matrix of $\boldsymbol Y$ with  descending order ${{\gamma_1} \ge  \ldots  \ge {\gamma_l}  > 0}$.
Then, ${\rm SPE}_{\scriptscriptstyle {OLPP}}$ is utilized to detect the abnormal change in the residual subspace and calculated as
\begin{center}
 $ {\rm SPE}_{\scriptscriptstyle {OLPP}} = {\left\| {\boldsymbol x - \widehat {\boldsymbol x}} \right\|^2}$,
\end{center}
where $\boldsymbol x$ is preprocessed by data normalization and $\widehat {\boldsymbol x}$ is the reconstruction of $\boldsymbol x$. The system model based on OLPP is given as
\begin{center}
  $ {\boldsymbol x}=   \widehat {\boldsymbol x} +{\boldsymbol e} ={\boldsymbol W}_{\scriptscriptstyle {OLPI}}   {\boldsymbol y}+{\boldsymbol e}$
\end{center}
for which (\ref{y-representation}) provides the optimal solution of minimizing $\|{\boldsymbol e}\|^2$,  then we have $\widehat{\boldsymbol x}=\boldsymbol{W}_{\scriptscriptstyle {OLPI}}  \boldsymbol {W}_{\scriptscriptstyle {OLPI}} ^{\rm T}  \boldsymbol{x}$.
 \begin{align}\label{eq8}
  {\rm SPE}_{\scriptscriptstyle {OLPP}}=&{{\boldsymbol x}}^{\rm T} \big(\boldsymbol{I}- \boldsymbol{W}_{\scriptscriptstyle {OLPI}} \boldsymbol{W}^{\rm T}_{\scriptscriptstyle {OLPI}}   \big)^2 {{\boldsymbol x}}\nonumber\\
  =  & {{\boldsymbol x}}^{\rm T} \big(\boldsymbol{I}- \boldsymbol{W}_{\scriptscriptstyle {OLPI}} \boldsymbol{W}^{\rm T}_{\scriptscriptstyle {OLPI}}   \big)  {{\boldsymbol x}}\nonumber\\
   = & \left\| \boldsymbol x \right\|^2 - \left\| \boldsymbol y \right\|^2
 \end{align}
 by making use of ${\boldsymbol W_{\scriptscriptstyle {OLPI}}^{\rm T}}{\boldsymbol W_{\scriptscriptstyle {OLPI}}}= {\boldsymbol I}$.

For OLPP-MLE, our proposed $T_{\scriptscriptstyle {OLPP}}^2$ and ${\rm SPE}_{\scriptscriptstyle {OLPP}}$ serve as significant indices to monitor the process, and the associated thresholds provide the reference to judge whether faults occur or not. Because the proposed OLPP-MLE is free from data distribution, kernel density estimation (KDE) technique is adopted to calculate thresholds, which is presented concisely below~\cite{kde}.

Given a random variable $ z$, $p(z)$ is the associated probability density function, then
\begin{equation}\label{p2}
 \mathop p\limits^ \wedge  ( z) = \frac{1}{{N \kappa}}\sum\limits_{n = 1}^N {\psi (\frac{{ z - { z_n}}}{\kappa })},
\end{equation}
where ${{z_n}(n = 1, \ldots ,N)}$ can represent ${T_{\scriptscriptstyle {OLPP}}^2}$ and ${\rm SPE}_{\scriptscriptstyle {OLPP }}$ statistics aforementioned. $\kappa $ is the bandwidth of kernel function $\psi \left(  \cdot  \right)$, which influences the estimation of $p(z)$ seriously.  In this paper, the optimal value $\kappa_{opt}$ is calculated by the following criterion, where $\sigma$ is standard  deviation~\cite{kde}:
\begin{equation}\label{hopt}
 {\kappa_{opt}} = 1.06\sigma {N^{ - 1/5}}.
\end{equation}
Given a confidence limit $\alpha$, the thresholds $J_{th,T^2}$ and $J_{th,{\rm SPE}}$ are computed by
 \begin{equation}\label{jt2}
   \int_{ - \infty }^{{J_{th,T^2}}} {\mathop p\limits^ \wedge  (T_{ {\scriptscriptstyle {OLPP}}}^2)d} {T_{\scriptscriptstyle  {OLPP}}^2} = \alpha,
 \end{equation}
 \begin{equation}\label{jspe2}
   \int_{ - \infty }^{{J_{th,{\rm SPE}}}} {\mathop p\limits^ \wedge  ({\rm SPE}_{\scriptscriptstyle {OLPP}})d} {{\rm SPE}_{\scriptscriptstyle {OLPP}}} = \alpha
 \end{equation}

\subsection{Summary of the proposed approach}
 The main procedure of OLPP-MLE fault diagnosis algorithm can be summarized thereafter.

 The off-line modeling phase is depicted as follows.

\begin{enumerate}[(1)]
 \item  Data normalization. Calculate the mean and standard deviation of the history data, which are normalized to zero mean and scaled to unit variance.

 \item  Estimate the ID of data via (\ref{mle-intrinsic}-\ref{estimatel}).

 \item Calculate the orthogonal locality preserving projections  ${\boldsymbol{W}_{\scriptscriptstyle {OLPI}}} $ according to SI file. 

 \item  Obtain the lower-dimensional representation $\textbf{\emph{y}}$ by (\ref{y-representation}).

 \item  Calculate the monitoring statistics $T_{\scriptscriptstyle {OLPP}}^2$ and ${{\rm SPE}_{\scriptscriptstyle {OLPP}}}$   according to (\ref{eq7}-\ref{eq8}).

 \item  Calculate the thresholds corresponding to the statistics aforementioned by KDE technique, described by (\ref{p2}-\ref{jspe2}).

 \end{enumerate}

The on-line monitoring phase is presented below.

 \begin{enumerate}[(1)]
 \item Collect and preprocess data. According to the mean and standard deviation in the off-line modeling phase, preprocess the collected testing data.

 \item Calculate the lower-dimensional representation of new data via (\ref{y-representation}).

 \item Calculate two monitoring statistics $T_{\scriptscriptstyle {OLPP}}^2$ and ${{\rm SPE}_{\scriptscriptstyle {OLPP}}}$ of new data according to (\ref{eq7}-\ref{eq8}).

 \item Judge whether a fault happens based on the fault detection logic:

  ${\rm SPE}_{\scriptscriptstyle {OLPP}}\le {J_{th,\rm SPE}}$ and $T_{\scriptscriptstyle {OLPP}}^2 \le {J_{th,T^2}}$ $ \Rightarrow $ fault free, otherwise faulty.
 \end{enumerate}

 Two indices are generally utilized to evaluate the algorithm accuracy, namely, fault detection rate (FDR) and false alarm rate (FAR), which could be calculated below.
\begin{equation}\label{mar}
\rm{FDR} = \frac{{number\;of\;samples\,\left( {\emph{J} >  {\emph{J}_{\emph{th}}}|\emph{f}\ne 0} \right)}}{{total\;samples\,\left( {\emph{f} \ne 0} \right)}}\times 100\%
\end{equation}
\begin{equation}\label{far}
\rm{FAR} = \frac{{number\;of\;samples\,\left( {\emph{J} > {\emph{J}_{\emph{th}}}|\emph{f} = 0} \right)}}{{total\;samples\,\left( {\emph{f} = 0} \right)}}\times 100\%
\end{equation}
where $J$ can be replaced by $T_{\scriptscriptstyle {OLPP}}^2$ or ${{\rm SPE}_{\scriptscriptstyle {OLPP}}}$, and $J_{th}$ is the associated threshold.


\subsection{Remarks}\label{remarks}
  \begin{itemize}
    \item  {{ \textbf{ \it Solutions of singular matrix $\boldsymbol X \boldsymbol D \boldsymbol X^{\rm T}$  }}}

   For the crucial step of calculating OLPP, matrix ${(\boldsymbol X \boldsymbol D \boldsymbol X^{\rm T})^{-1}}$ is considerably critical. However, $\boldsymbol X \boldsymbol D \boldsymbol X^{\rm T}$ is singular in most cases, especially when $\boldsymbol X$ has data redundancy or is normalized to zero mean. Thus, the inverse matrix $(\boldsymbol X \boldsymbol D \boldsymbol X^{\rm T})^{-1}$ does not actually exist. In our approach, three alternative methods are provided to cope with this issue.

   1) {PCA projection}

Before conventional OLPP, $\boldsymbol x$ is firstly projected into PCA subspace and then irrelevant features are extracted. Let $\boldsymbol W_{\scriptscriptstyle {PCA}}$ denote the transformation matrix via PCA and $\widehat{\boldsymbol X}$ denote the reconstruction of $\boldsymbol X$ after PCA projection.
As the preprocessing step of OLPP, it ensures that matrix ${\widehat{\boldsymbol X} {\boldsymbol D} \widehat {\boldsymbol X}^{\rm T}}$ is nonsingular. In the following procedure, ${\boldsymbol X}$ is replaced by $\widehat{\boldsymbol X}$ accordingly.  At the fourth step of OLPP in SI file, the transformation matrix $\boldsymbol W$ is calculated by
\begin{equation}\label{eq6}
 \boldsymbol W=\boldsymbol W_{\scriptscriptstyle {PCA}}{\boldsymbol W_{\scriptscriptstyle {OLPI}}}.
\end{equation}
Since vectors in $\boldsymbol W_{\scriptscriptstyle {PCA}}$ and $\boldsymbol W_{\scriptscriptstyle {OLPI}}$  are orthonormal, vectors in $\boldsymbol  W$ are still mutually orthonormal. Hence, $\boldsymbol W$  should be substituted for $\boldsymbol W_{\scriptscriptstyle {OLPI}}$  in (\ref{y-representation}), and the rest procedure remains the same.

2) {Regularization}

Regularization is a common technique to cope with singular problem.
In the proposed approach, the regularization term is utilized by adding constant values to the diagonal elements of ${\boldsymbol X \boldsymbol D \boldsymbol X^{\rm T}}$, as ${{\boldsymbol X \boldsymbol D \boldsymbol X^{\rm T}}+\beta \boldsymbol I}$, where $\beta>0$ is predefined by users. It is obviously that ${\boldsymbol X\boldsymbol D \boldsymbol X^{\rm T}+\beta \boldsymbol I}$ is non-singular.

3) {Pseudo inverse}

When ${\boldsymbol X \boldsymbol D \boldsymbol X^ {\rm T}} $ is singular, the concept of pseudo inverse is introduced, which denotes as $(\cdot)^{\dagger}$. Suppose ${rank(\boldsymbol X \boldsymbol D \boldsymbol X^ {\rm T})= r}$, the singular value decomposition (SVD) of this matrix is
\begin{equation}
  \boldsymbol X \boldsymbol D \boldsymbol X^{\rm T} = \boldsymbol P {\boldsymbol \Sigma} {\boldsymbol Q^{\rm T}},
\end{equation}
where ${{\boldsymbol \Sigma} \in {\mathcal R}^{r\times r}}$, ${{\boldsymbol P, \boldsymbol Q} \in {\mathcal R}^{N\times r}}$, $ \boldsymbol P^{\rm T} {\boldsymbol P}= \boldsymbol Q^{\rm T} {\boldsymbol Q} = \boldsymbol I$.
Then, pseudo inverse of matrix ${\boldsymbol X \boldsymbol D \boldsymbol X^{\rm T}}$ can be computed by
\begin{equation}
  {\left( {\boldsymbol X \boldsymbol D \boldsymbol X^ {\rm T}} \right)^{\dagger }} =\boldsymbol Q{ \boldsymbol \Sigma ^{ - 1}}{\boldsymbol P^{\rm T}}.
\end{equation}
Thus,  ${\left( {\boldsymbol X \boldsymbol D \boldsymbol X^ {\rm T}} \right)^{\dagger }}$ is substituted for ${\left( {\boldsymbol X \boldsymbol D \boldsymbol X^ {\rm T}} \right)^{-1 }}$ in the following derivation process.

 \item  {\it Another perspective to calculate vector $\boldsymbol a_1$}

   With regard to vector $\boldsymbol a_1$, it is the eigen-vector of $\boldsymbol X \boldsymbol L \boldsymbol X^{\rm T} \boldsymbol a =  \lambda \boldsymbol X \boldsymbol D \boldsymbol X^{\rm T} \boldsymbol a$ corresponding to the smallest eigenvalue. Three alternative solutions aforementioned are fairly effective when $\boldsymbol X \boldsymbol D \boldsymbol X^{\rm T}$ is singular. Besides, SVD approach can  also be widely employed to settle the underlying singular problem~\cite{eigenvector_svd}.

Suppose $rank(\boldsymbol X) = r_x$, the SVD of $\boldsymbol X$ is
\begin{equation} 
 \boldsymbol X =\boldsymbol U {\boldsymbol \Sigma_x} {\boldsymbol V^{\rm T}}
\end{equation}
where ${{\boldsymbol \Sigma_x} \in {\mathcal R}^{r_x \times r_x}} $, ${\boldsymbol U,\boldsymbol V \in {\mathcal R^{N \times {r_x}}}}$ and ${\boldsymbol U^{\rm T}}{\boldsymbol U}={\boldsymbol V^{\rm T}}{\boldsymbol V}=\boldsymbol I$.
Let $\boldsymbol b = \boldsymbol \Sigma_x {\boldsymbol U^{\rm T}} {\boldsymbol a}$, then
\begin{equation}
\begin{aligned}
                 &  \boldsymbol X \boldsymbol L{\boldsymbol X^{\rm T}} \boldsymbol a = \lambda \boldsymbol X \boldsymbol D{\boldsymbol X^T}\boldsymbol a \\
\Rightarrow \;\; &  \boldsymbol U \boldsymbol \Sigma_x {\boldsymbol V^{\rm T}}\boldsymbol L \boldsymbol V \boldsymbol \Sigma_x {\boldsymbol U^{\rm T}}\boldsymbol a = \lambda \boldsymbol U \boldsymbol \Sigma_x {\boldsymbol V^{\rm T}}\boldsymbol D \boldsymbol V \boldsymbol \Sigma_x {\boldsymbol U^{\rm T}}\boldsymbol a\\
\Rightarrow \;\; &  \boldsymbol U \boldsymbol \Sigma_x {\boldsymbol V^{\rm T}}\boldsymbol L \boldsymbol V \boldsymbol b = \lambda \boldsymbol U \boldsymbol \Sigma_x {\boldsymbol V^{\rm T}}\boldsymbol D \boldsymbol V \boldsymbol b \\
\Rightarrow \;\; &  {\boldsymbol \Sigma_x ^{ - 1}}{\boldsymbol U^{\rm T}}\boldsymbol U \boldsymbol \Sigma_x {\boldsymbol V^{\rm T}}\boldsymbol L \boldsymbol V \boldsymbol b = \lambda {\boldsymbol \Sigma_x ^{ - 1}}{\boldsymbol U^{\rm T}}\boldsymbol U \boldsymbol \Sigma_x {\boldsymbol V^{\rm T}}\boldsymbol D \boldsymbol V \boldsymbol b \\
\Rightarrow \;\; &  {\boldsymbol V^{\rm T}} \boldsymbol L \boldsymbol V \boldsymbol b = \lambda {\boldsymbol V^{\rm T}}\boldsymbol D \boldsymbol V \boldsymbol b. \label{solutionb}
\end{aligned}
\end{equation}
It is obvious that $\boldsymbol V^{\rm T} \boldsymbol D \boldsymbol V$ is nonsingular and the generalized eigen-problem in (\ref{solutionb}) can be easily obtained. After $\boldsymbol b^*$ is obtained, $\boldsymbol a^*$ is computed by
\begin{equation}
  \boldsymbol a^* = \boldsymbol U{\boldsymbol \Sigma_x ^{ - 1}}\boldsymbol b^*.
\end{equation}

     \item  {\it Computational complexity analysis}:
According to the procedure of OLPP-MLE, the computation mainly contains  adjacency graph construction, embedding functions and the estimation of ID via MLE. Since the computational complexity of $k$-nearest neighbor (KNN) is $O(N)$,  thus the computational complexity of adjacency graph construction and MLE is also $O(N)$, which both obtain excellent relevant results based on KNN. The embedding functions needs  $l$ times SVD on $m\times m$ matrix and $4(l-1)$ times matrix inversion on $m\times m$ matrix.

 \end{itemize}

\section{Numerical case and CSTR case studies}\label{numerical-case}

\subsection{Numerical example}
Considering the following system
\begin{center}
	$\left\{ \begin{array}{l}
	{x_1} = t + {\varepsilon _1}\\
	{x_2} = \cos t + {\varepsilon _2}\\
	{x_3} = {t^2} + t + {\varepsilon _3}
	\end{array} \right.,$
\end{center}
where ${t \in [ - 1,1]}$ and noise term ${\varepsilon_i (i = 1,2,3)}$ follow uniform distribution with ${\varepsilon _i} \in [ - 0.05,0.05]$.

$1000$ normal samples are generated to train the process monitoring model and then artificial faults are generated with $1000$ samples by the following scheme:

1) Fault $1$:  variable $x_1$ is added by 0.6 from the $501$th sample;

2) Fault $2$:  variable $x_2$ is added by 0.8 from the $501$th sample;

3) Fault $3$:  variable $x_3$ is added by 1.0 from the $501$th sample.

Note that $T_{\scriptscriptstyle {OLPP}}^2$ and ${{\rm SPE}_{\scriptscriptstyle {OLPP}}}$ are recorded briefly as $T^2$ and SPE in simulation figures for simplicity and convenience.

\begin{figure*}[!htp]
\centering
\subfigure[Fault 1]{
\label{numerical1}
\includegraphics[width=0.315\textwidth,height=4.5cm,angle=0]{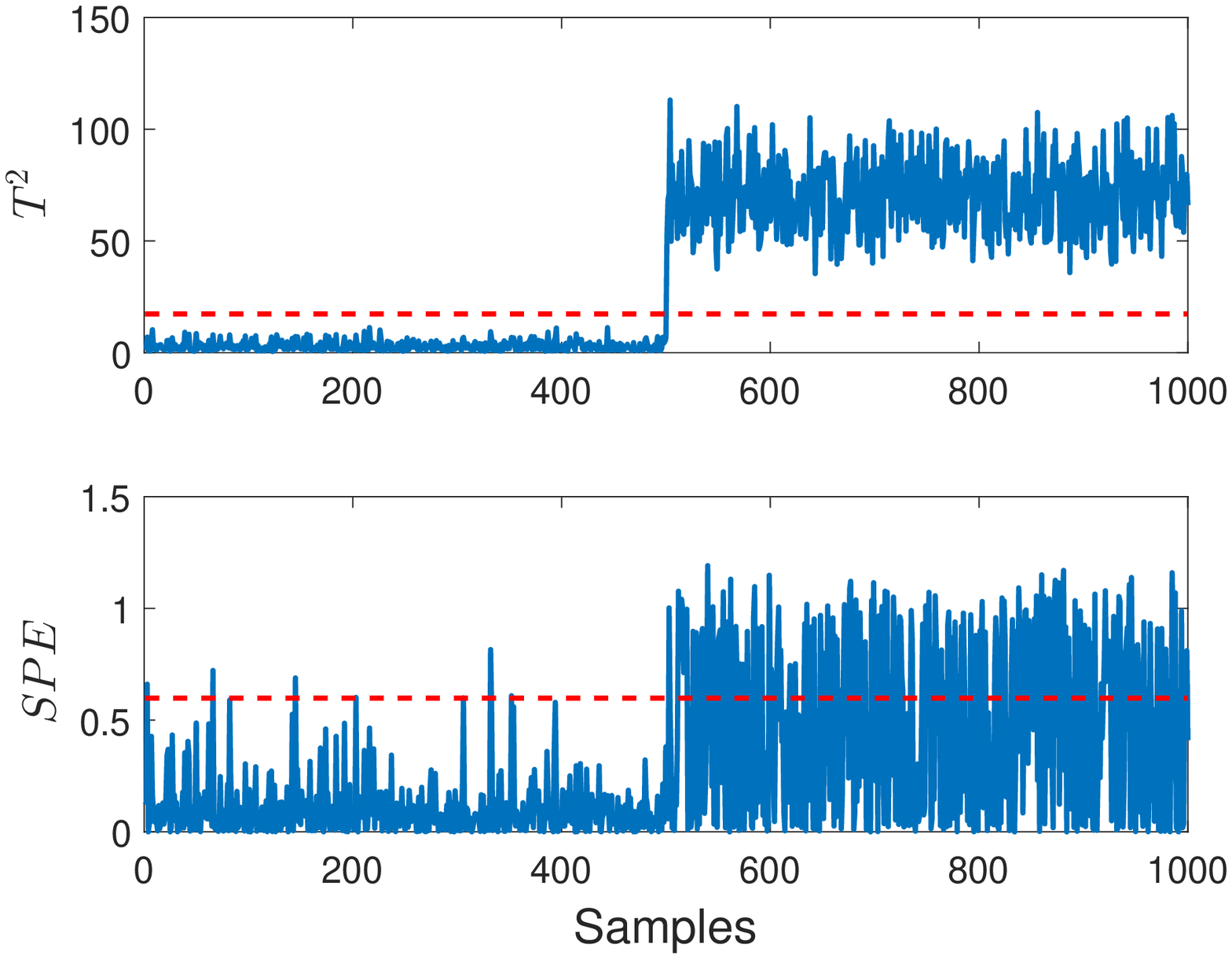}
}
\hspace{-2mm}
\subfigure[Fault 2]{
\label{numerical2}
\includegraphics[width=0.315\textwidth,height=4.5cm,angle=0]{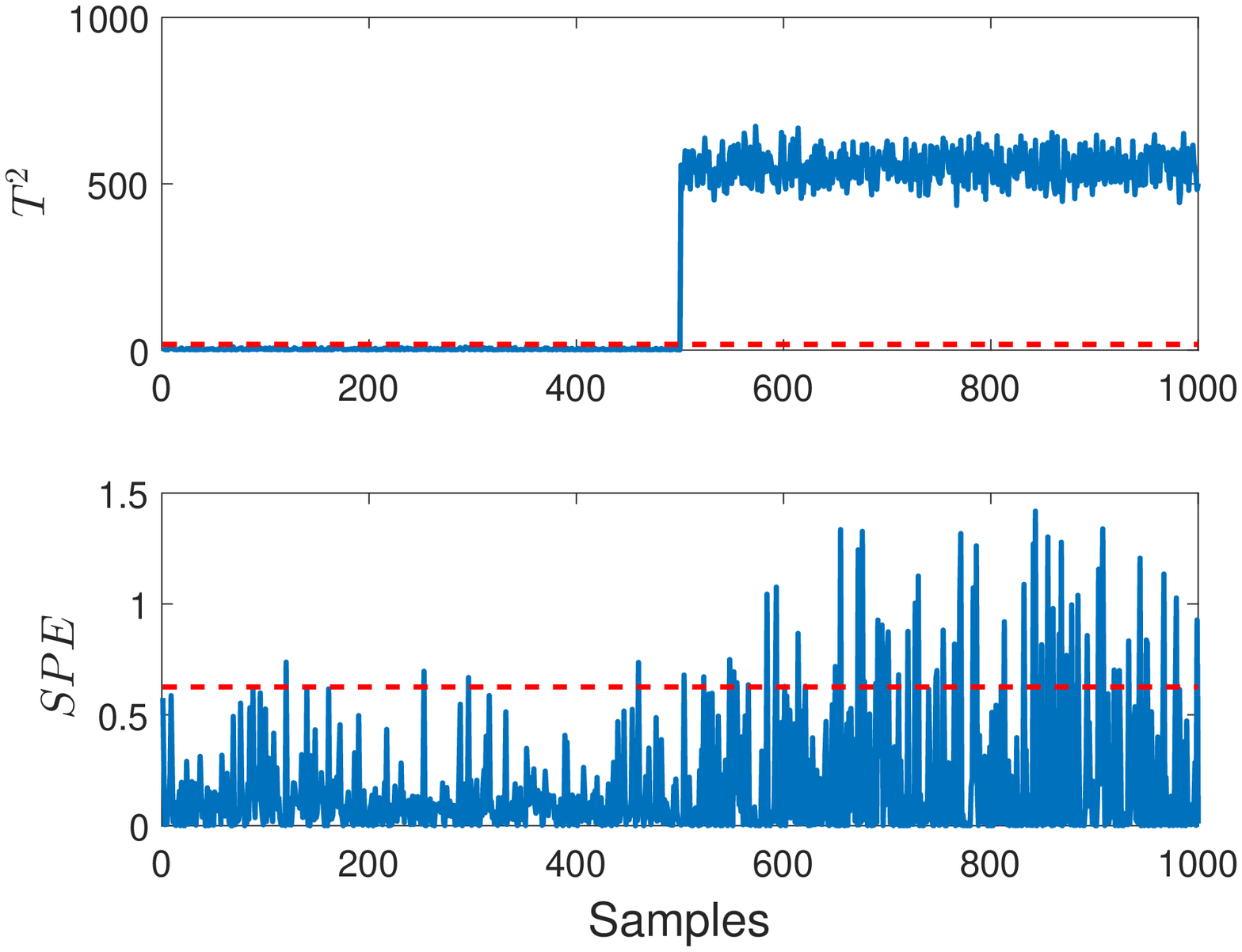}
}
\hspace{-2mm}
\subfigure[Fault 3]{
\label{numerical3}
\includegraphics[width=0.315\textwidth,height=4.5cm,angle=0]{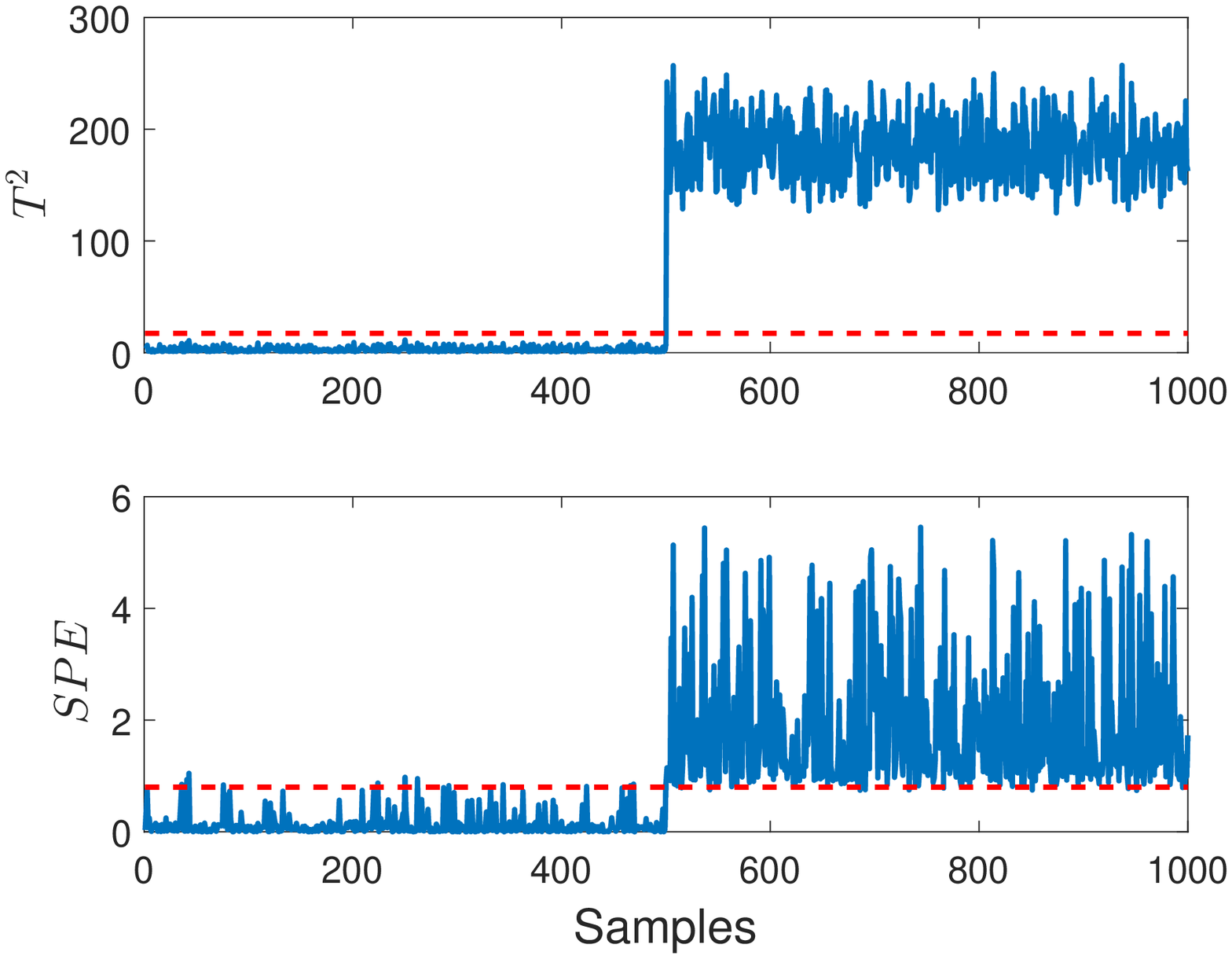}
}
\caption{Monitoring charts of numerical case}
\label{numerical-simulation}
\end{figure*}

Monitoring consequences are illustrated in Figure~\ref{numerical-simulation}. The FDRs of three faults nearly approach to $100\%$ and the FARs are all below $5\%$. Specifically, with regard to Faults $1$ and $2$, $T^2$ monitoring statistic can totally detect the fault while SPE has several missed alarm points. As illustrated in Figure \ref{numerical3}, both monitoring statistics can detect Fault $3$ timely and accurately.
In conclusion, faults can be detected by the cooperation of $T^2$ and SPE monitoring statistics, which indicates that OLPP-MLE is able to monitor this process.
\subsection{Case study on CSTR}
 In this section,  we employ MLE to estimate the ID, then PCA, LPP and OLPP are adopted to monitor the process. Thus, PCA-MLE, LPP-MLE and OLPP-MLE are compared and the superiority of OLPP-MLE is illustrated  by CSTR.

\subsubsection{CSTR introduction}
The dynamic behavior of CSTR process is depicted as follows \cite{shang2015concurrent}:
\begin{equation}
 \frac{{d{C_A}}}{{dt}} = \frac{q}{V}\left( {{C_{Af}} - {C_A}} \right) - {k_0}\exp \left\{ { - \frac{E}{{RT}}} \right\}{C_A} + {v_1}
\end{equation}
\begin{equation}
 \frac{{dT}}{{dt}} = \frac{q}{V}\left( {{T_f} - T} \right) - \frac{{\Delta H}}{{\rho {C_p}}}{k_0}\exp \left\{ { - \frac{E}{{RT}}} \right\}{C_A} + \frac{{UA}}{{V\rho {C_p}}}\left( {{T_c} - T} \right) + {v_2}
\end{equation}
where the outlet concentration $C_A$ and the outlet temperature $T$ are controlled by PI controllers, $q$ is the feed flow rate, $C_{Af}$ is the feed concentration, $V$ is the volume of the vessel, $T_f$ is the feed temperature, $v_1$ and $v_2$ are independent system noises \cite{ji2016incipient}.

In this simulation, the sampling interval is 1 second. The measured process variable $\left[ {{C_A}\,\;T\,\;{T_c}\,\;q} \right]\;$ are collected, and the measurement noise $e$ is added. Besides, negative feedback inputs were added to $\left[ {\;q\,\;{T_c}} \right]\;$ with PID controllers as ${\boldsymbol K_2}\left( {{K_1} + {{{T_d}s + {T_I}} \mathord{\left/  {\vphantom {{{T_d}s + {T_I}} s}} \right.
 \kern-\nulldelimiterspace} s}} \right)\varepsilon $, where $\varepsilon  = [{C_{Af}} - C_A^ * ,T - {T^ * }] $ is the residual vector. All system parameters and conditions are set as the same with Li et al \cite{li2010reconstruction}. 

\subsubsection{Monitoring results of CSTR case}
In this paper, 6000 normal samples are collected to establish the monitoring model, that is, PCA-MLE, LPP-MLE and OLPP-MLE. We collect samples of 600 minutes and the faults are designed as follows:

1) Fault 4: the feed temperature $T_f$ is increased by $1\%$ at the 101th min;

2) Fault 5: the volume of vessel $V$ is decreased at the rate of $\frac{{\rm{4}}}{{{\rm{500}}}}$ $m^3$/mins.

\begin{figure*}[!htbp]
\centering
\subfigure[PCA-MLE]{
\label{cstrpca1}
\includegraphics[width=0.315\textwidth,height=4.5cm,angle=0]{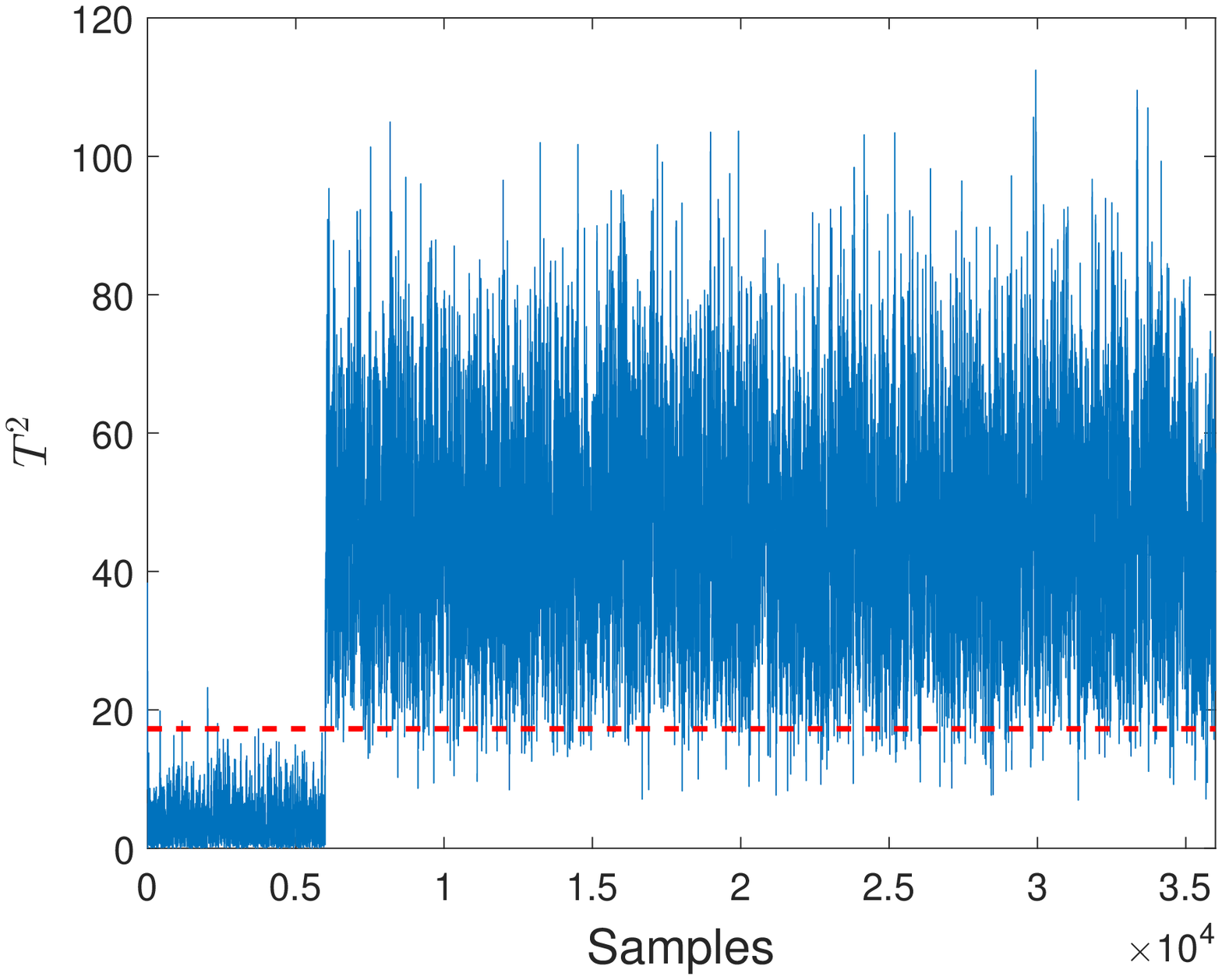}
}
\hspace{-2mm}
\subfigure[LPP-MLE]{
\label{cstrlpp1}
\includegraphics[width=0.315\textwidth,height=4.5cm,angle=0]{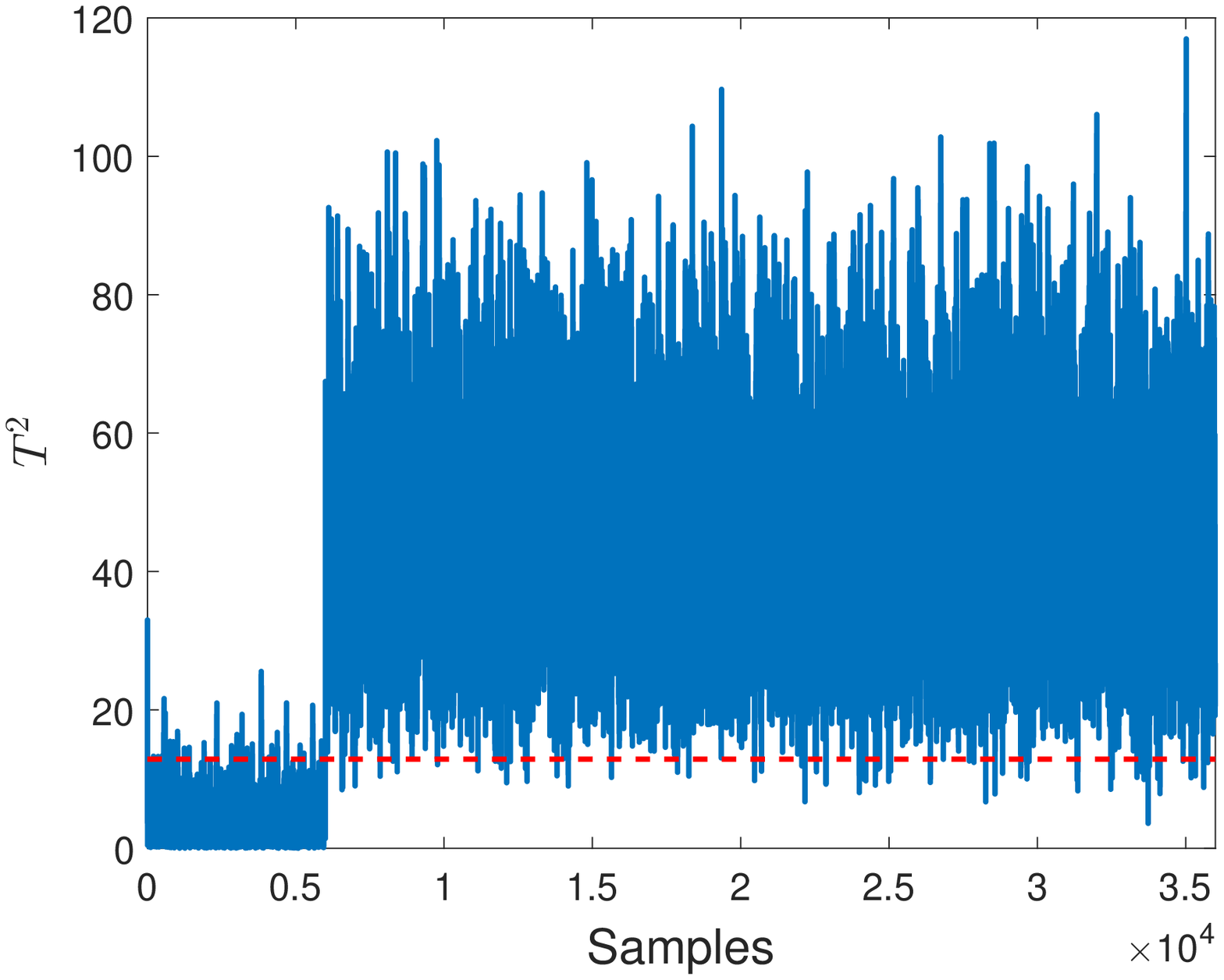}
}
\hspace{-2mm}
\subfigure[OLPP-MLE]{
\label{cstrolpp1}
\includegraphics[width=0.315\textwidth,height=4.5cm,angle=0]{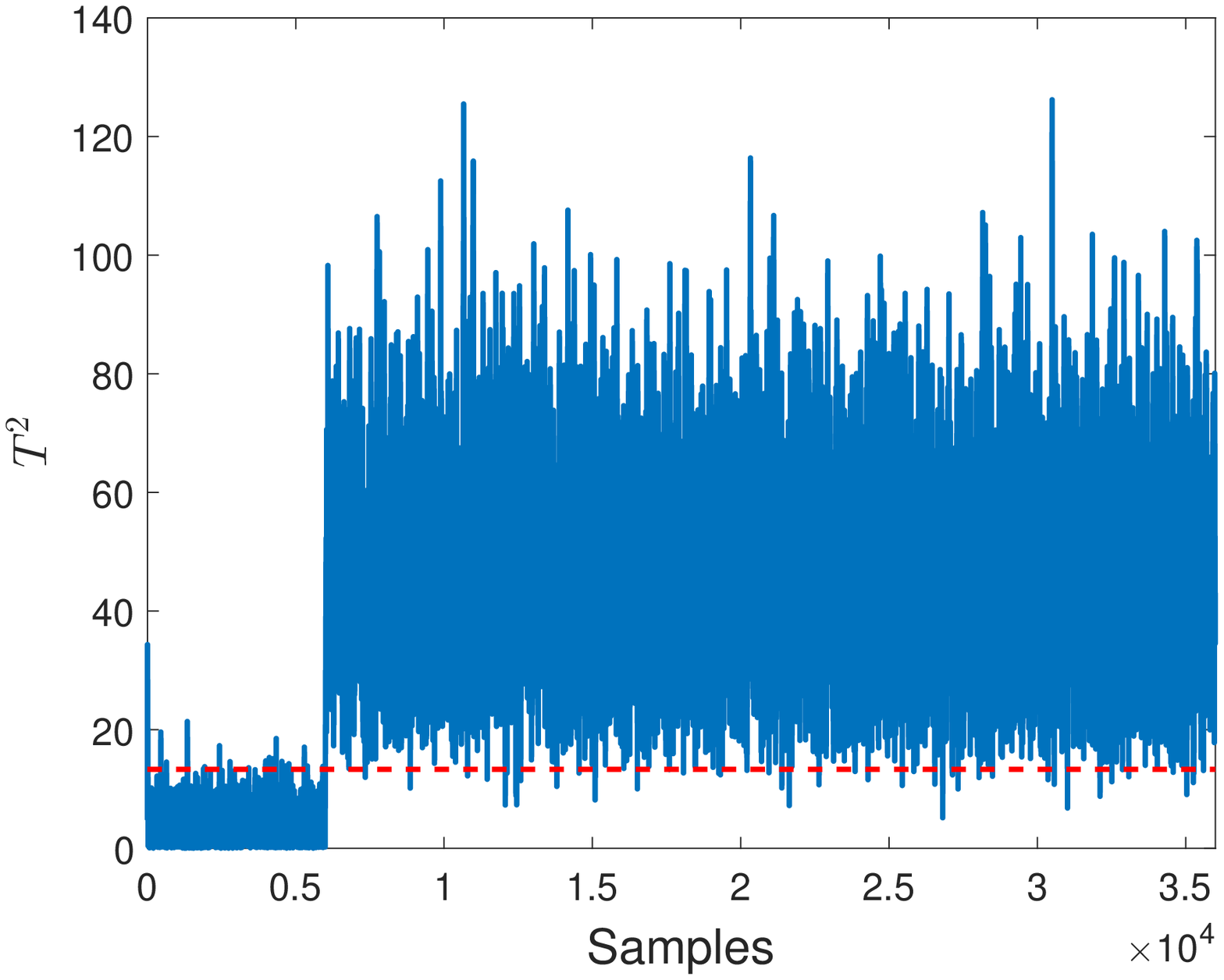}
}
\caption{Monitoring charts of Fault 4}
\label{fault4}
\end{figure*}

\begin{figure*}[!htbp]
\centering
\subfigure[PCA-MLE]{
\label{cstrpca3}
\includegraphics[width=0.315\textwidth,height=4.5cm,angle=0]{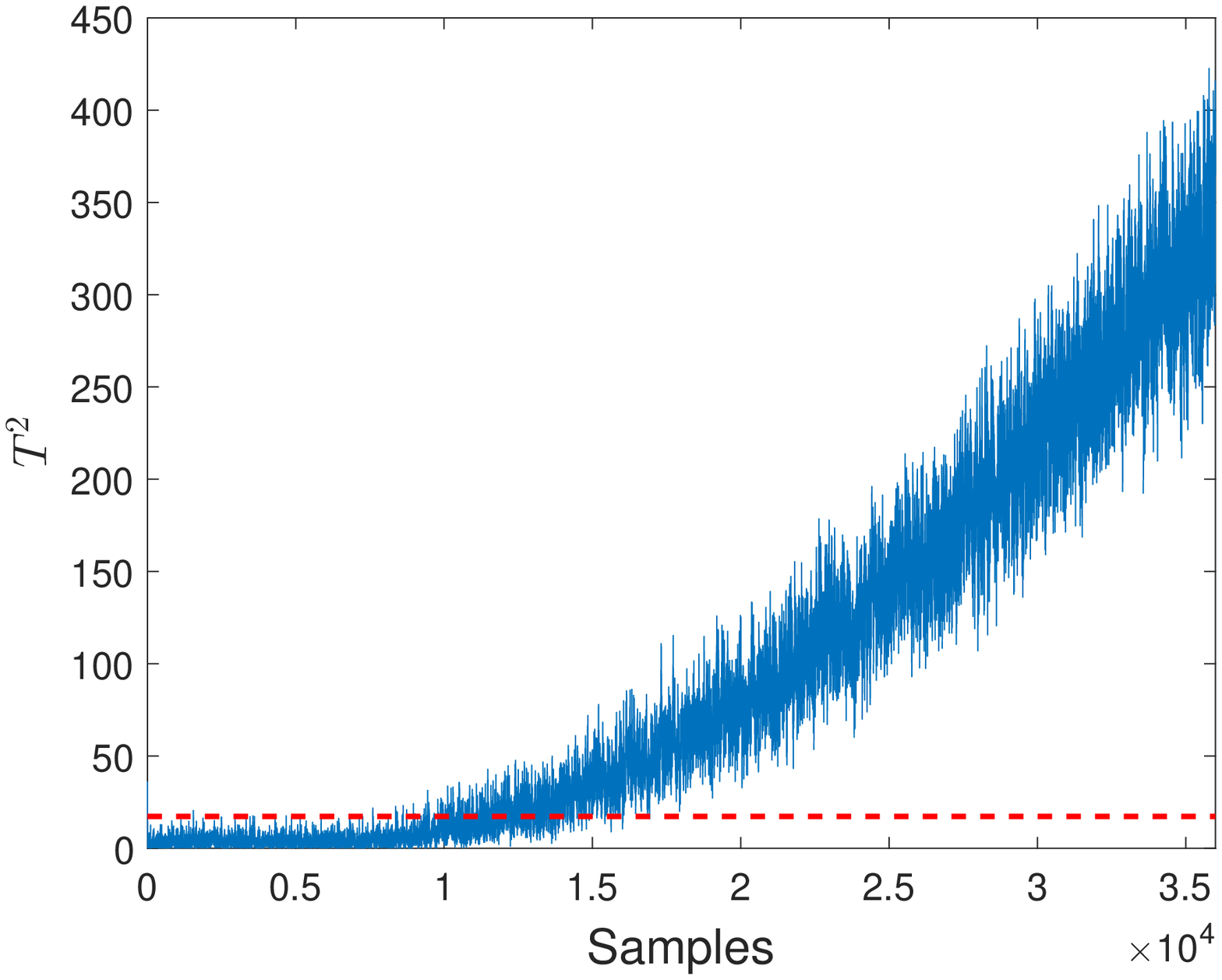}
}
\hspace{-2mm}
\subfigure[LPP-MLE]{
\label{cstrlpp3}
\includegraphics[width=0.315\textwidth,height=4.5cm,angle=0]{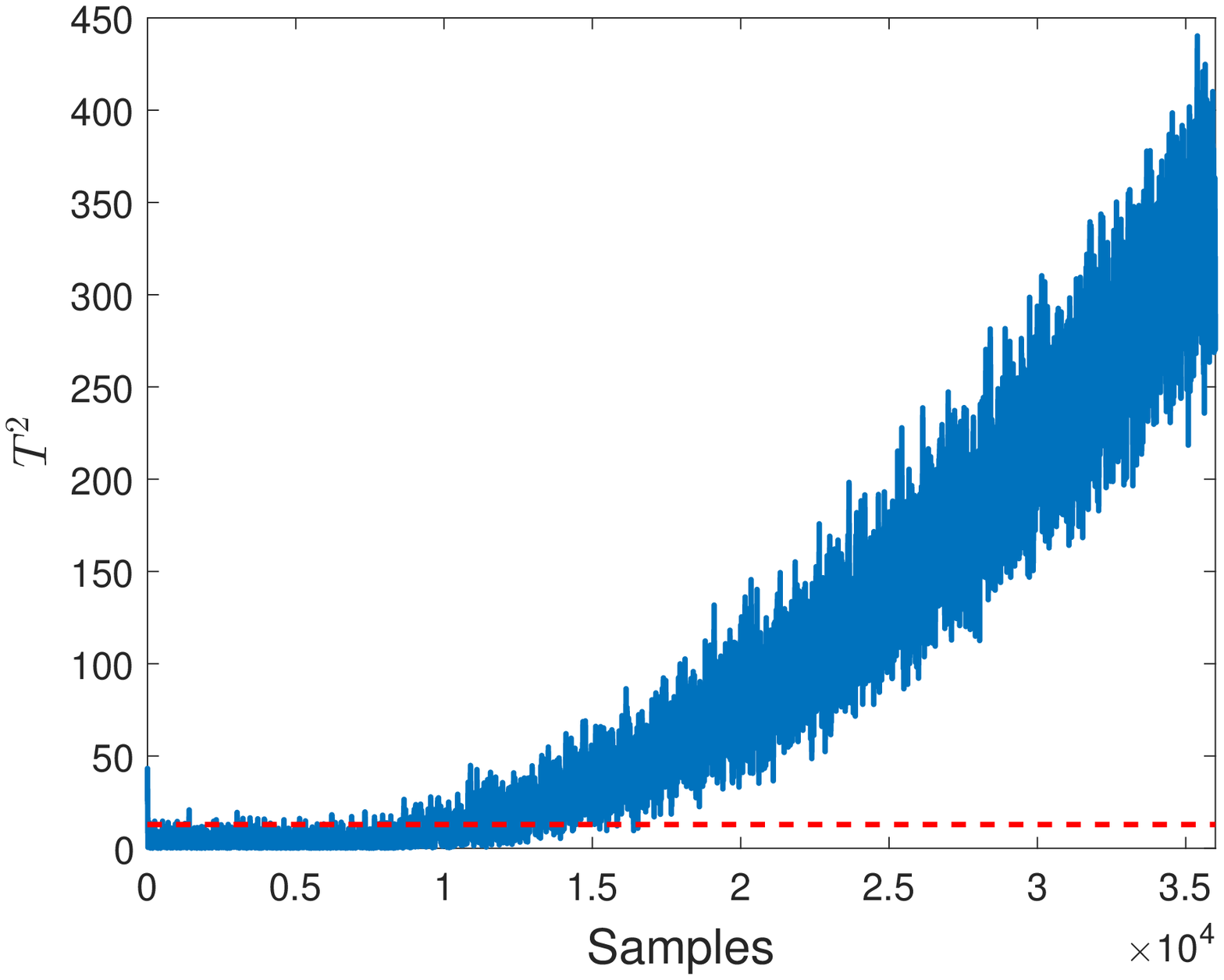}
}
\hspace{-2mm}
\subfigure[OLPP-MLE]{
\label{cstrolpp3}
\includegraphics[width=0.315\textwidth,height=4.5cm,angle=0]{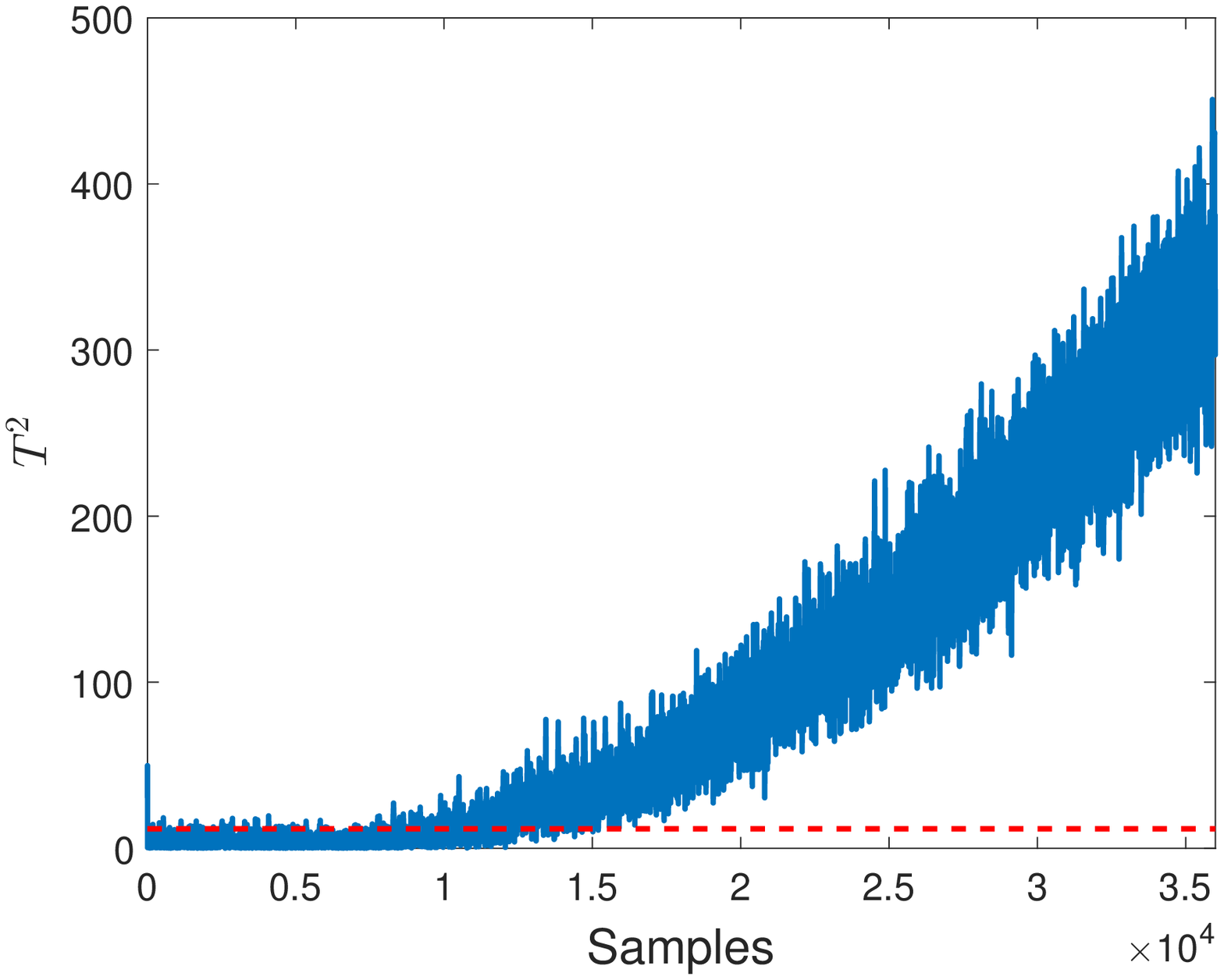}
}
\caption{Monitoring charts of Fault 5}
\label{fault5}
\end{figure*}

In this simulation, the raw data is preprocessed by a low pass filter to reduce noise. Then, PCA-MLE, LPP-MLE and OLPP-MLE are adopted to monitor the process. The intrinsic dimensionality is 4 through MLE technique. Thus, only $T^2$ monitoring statistic works. As exhibited in Figures \ref{fault4} and \ref{fault5}, it reveals that outlines of monitoring charts are similar for different approaches. The monitoring results are summarized in Table \ref{Table-cstr-results}. It shows that the FDRs of the proposed OLPP-MLE approach are the highest, especially for Fault 5. Moreover, the FARs are lower than $1\% $.

\begin{table*}[htp]
\renewcommand{\arraystretch}{1}
\caption{FDRs (\%) and FARs (\%) of CSTR case}
\centering
\begin{tabular}{c c c c}
\toprule
 Fault type         & PCA-MLE  & LPP-MLE    & OLPP-MLE \\
 \hline
 Normal operation   & 0.13      & 0.57      &0.50\\
 Fault 4            & 98.98     & 99.70     &99.80\\
 Fault 5            &78.85      & 82.20     &83.19\\
\bottomrule
\end{tabular}\label{Table-cstr-results}
\end{table*}

\section{Benchmark simulation and comparative study}\label{section5}
In this section, Tennessee Eastman process data is employed to prove the proposed approach. Besides, several existing data-driven approaches are utilized to compare with OLPP-MLE.

\subsection{Tennessee Eastman process}

\begin{figure}[!tp]
 \centering
\includegraphics[width=0.42\textwidth,angle=90]{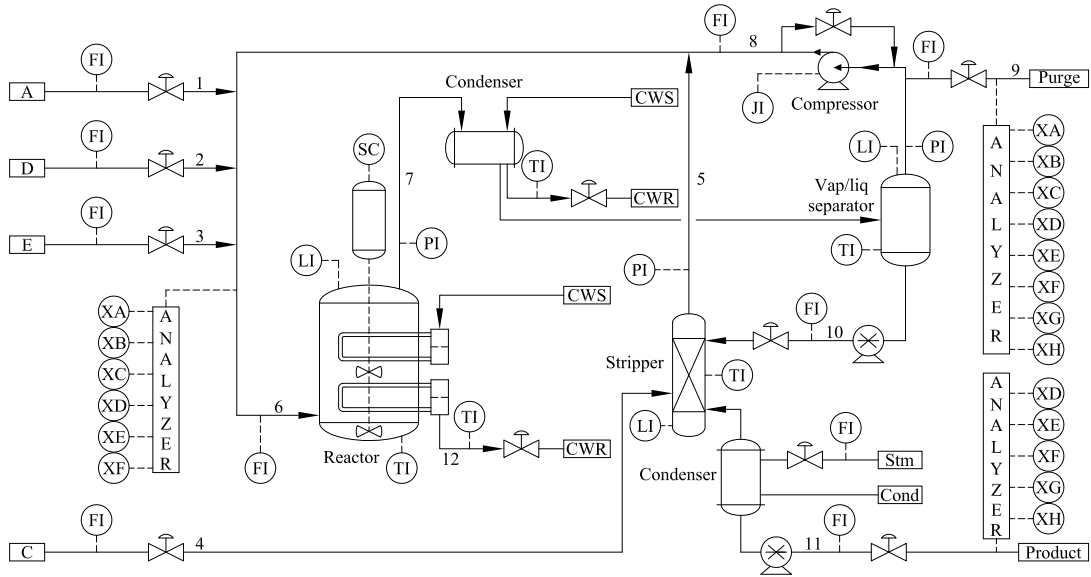}
\caption{The flow diagram of TE process}
\label{fig-te}
\end{figure}

TE process is a well-established simulator that is generally served as a preferred benchmark for fault detection research~\cite{TE2,comparison,LWPR}. The flow diagram is shown in Figure~\ref{fig-te} and more detailed information can be found in Down et al~\cite{TEP}.

$20$ process faults  and another valve fault were defined, namely, IDV(1)-IDV(21). IDV(0) represents normal operation condition. The types of faults include step, random variation, show drift, sticking, constant position and unknown faults~\cite{comparison}. Due to the frequent absence of sufficient process knowledge, it is necessary to employ data-driven techniques for process monitoring.

In this simulation,  $22$ process variables and $11$ manipulated variables are selected as the samples. $960$ normal samples are used to establish the off-line model. Then, $960$ testing samples, including the first $160$ normal samples and the following $800$ faulty samples, are utilized to evaluate the algorithm. Let $\alpha=0.99$ be the confidence level.

\subsection{Intrinsic dimensionality}\label{discussionofl}
\begin{figure*}[!htbp]
\centering
\subfigure[The range of $k$ ]{
\label{fig3}
\includegraphics[width=0.47\textwidth,angle=0]{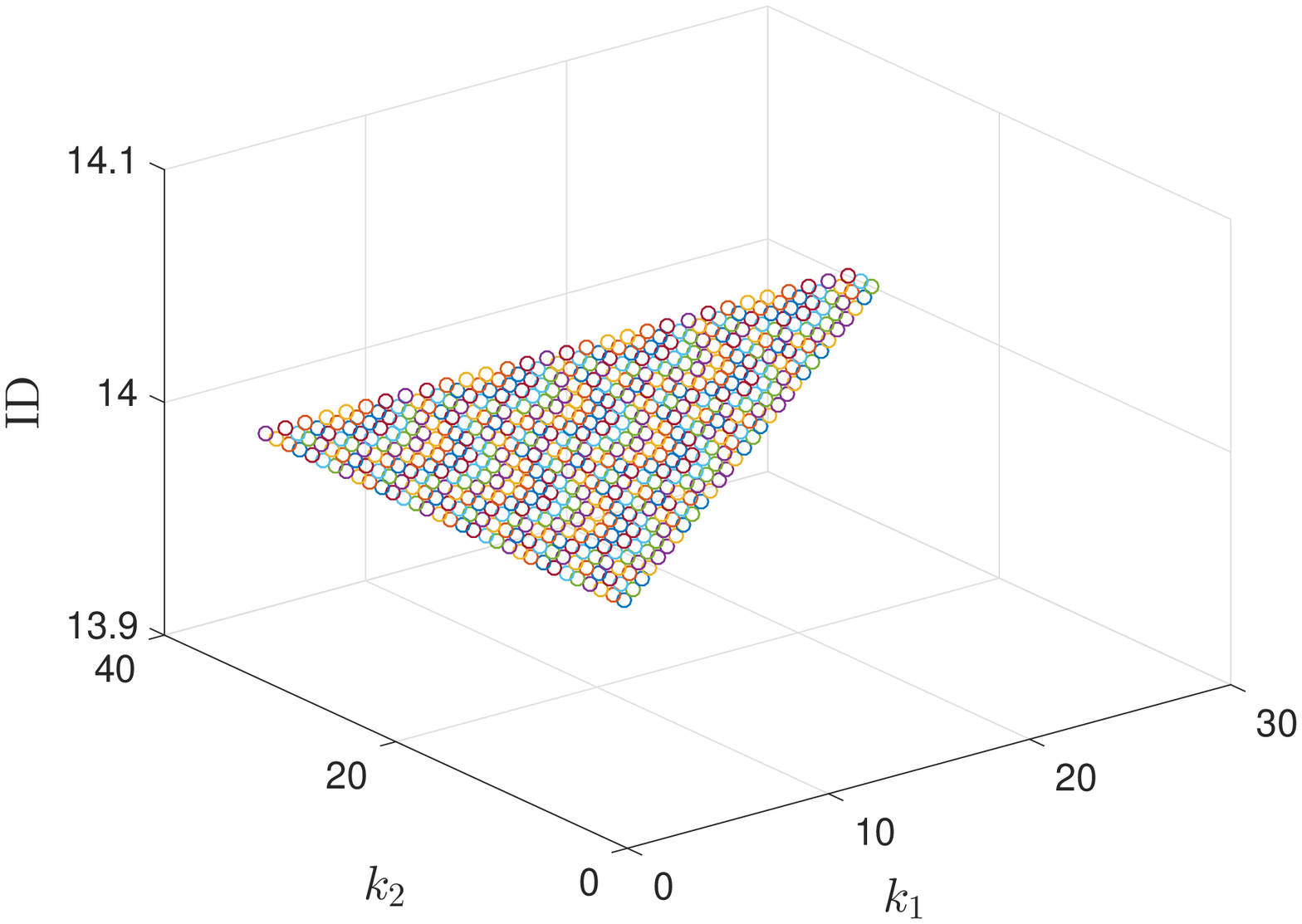}
}
\subfigure[Sampling frequency]{
\label{fig4}
\includegraphics[width=0.47\textwidth,angle=0]{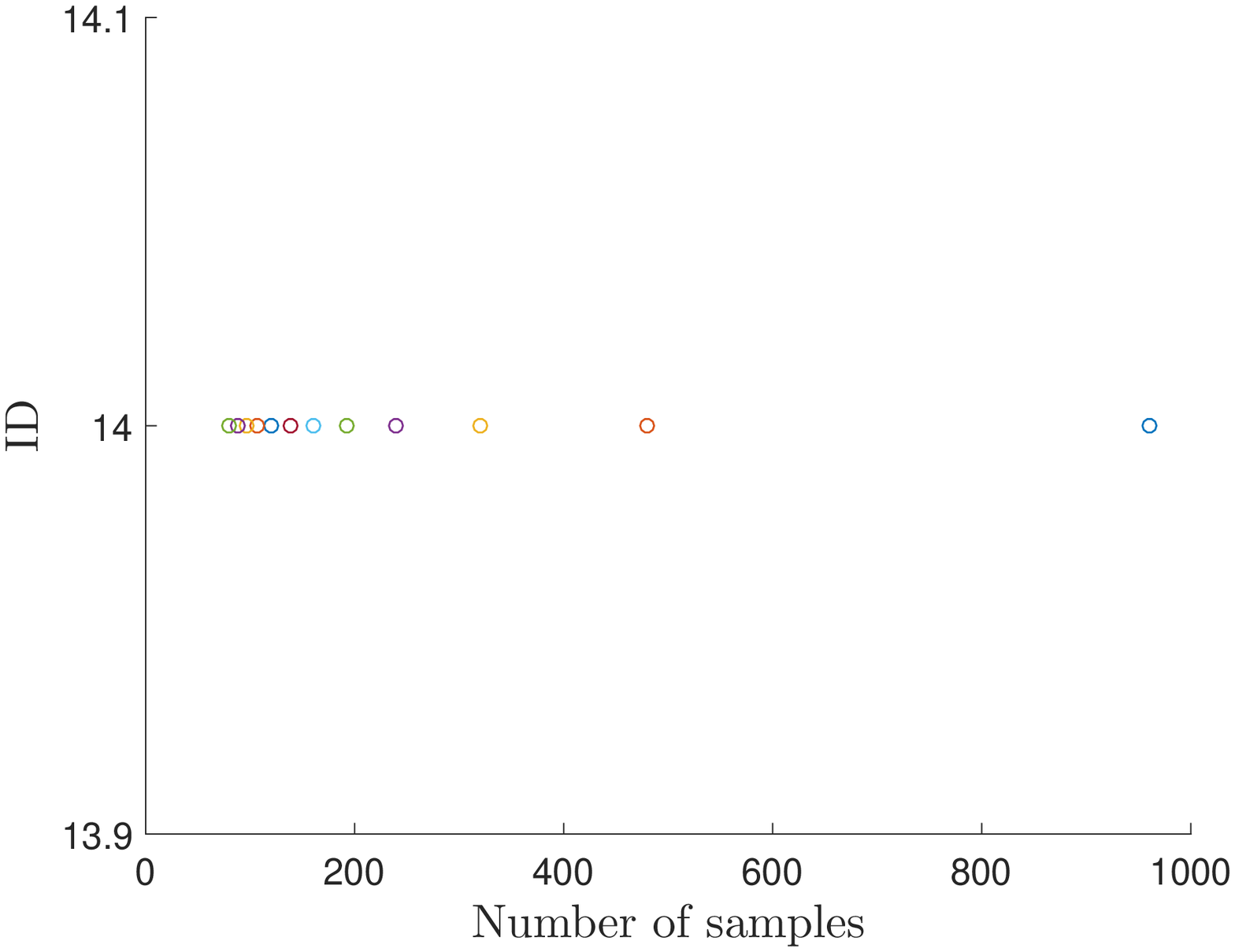}
}
\caption{The stability of estimating ID}
\label{TE-simulation1}
\end{figure*}

In this section, the parameter of MLE and the sampling frequency are discussed when the ID is estimated.  The influence of the range $k_1,\ldots,k_2$ is illustrated in Figure~\ref{fig3}, where $k_1 \in \left[ {1,30} \right]$ and $k_2 \in \left[ {k_1+1,33} \right]$.
The ID remains the same, which indicates that the estimation of ID is insensitive to the range of $k_1$, $k_2$.

Figure~\ref{fig4} demonstrates the influence of sampling frequency. As we have 960 normal data in total, every $j$ interval, samples are taken to acquire the ID. It denotes that the sampling frequency reduces to $1/j$ with $j\in \left[ {1,12} \right]$.  It can be discovered evidently that the ID keeps basically constant in Figure~\ref{fig4}. In summary, the estimation of ID is stable via MLE.

%

\subsection{Simulation results of OLPP-MLE}
Three typical faults, i.e., step, random variation and sticking,  are selected to demonstrate OLPP-MLE algorithm. Specifically, fault IDV(1) is utilized to illustrate the step fault, fault IDV(12) is used to account for random variation fault, and fault IDV(14) is employed to represent the sticking fault.

\begin{figure*}[!htbp]
\centering
\subfigure[IDV1]{
\label{IDV1}
\includegraphics[width=0.315\textwidth,height=4.5cm,angle=0]{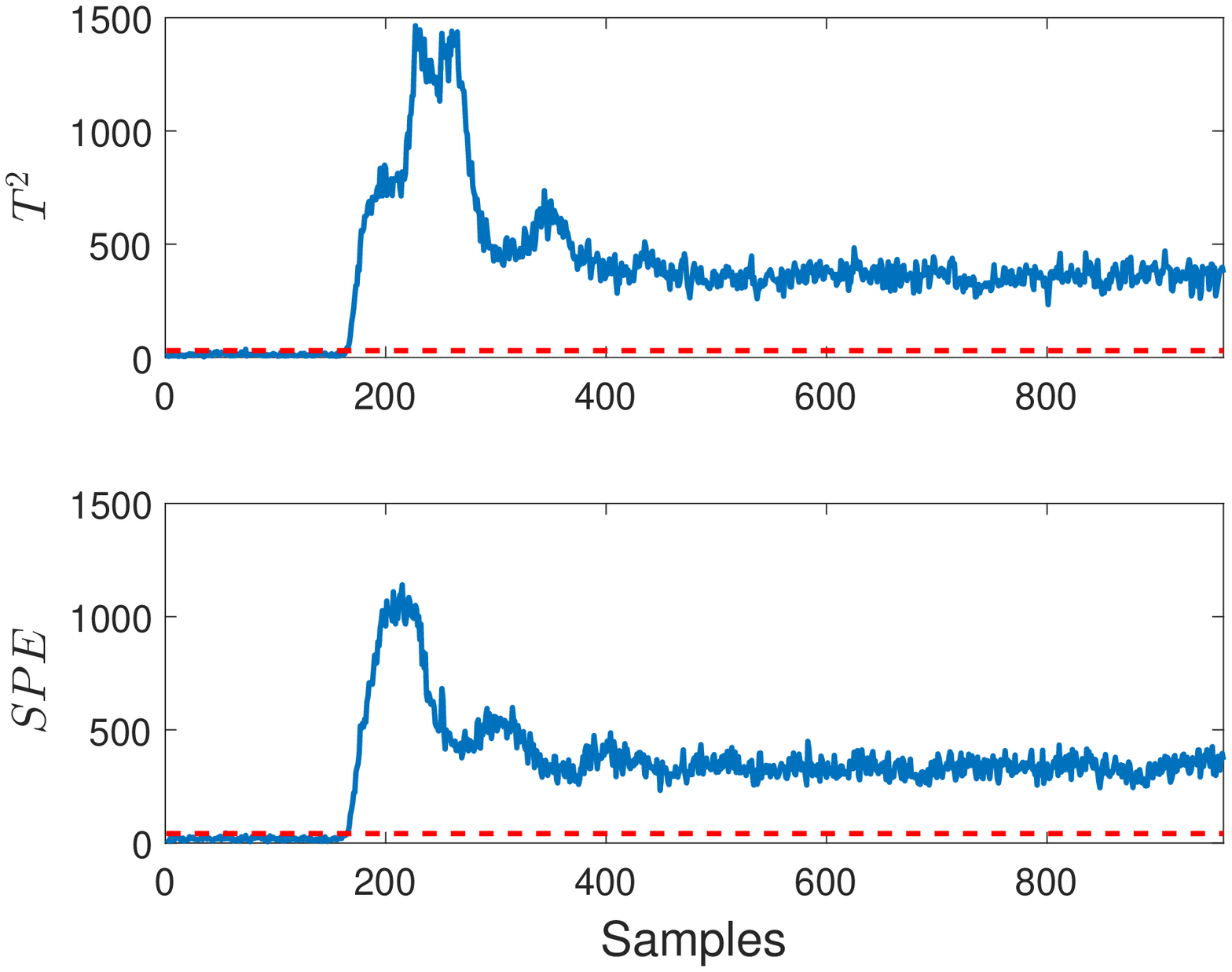}
}
\hspace{-2mm}
\subfigure[IDV12]{
\label{IDV12}
\includegraphics[width=0.315\textwidth,height=4.5cm,angle=0]{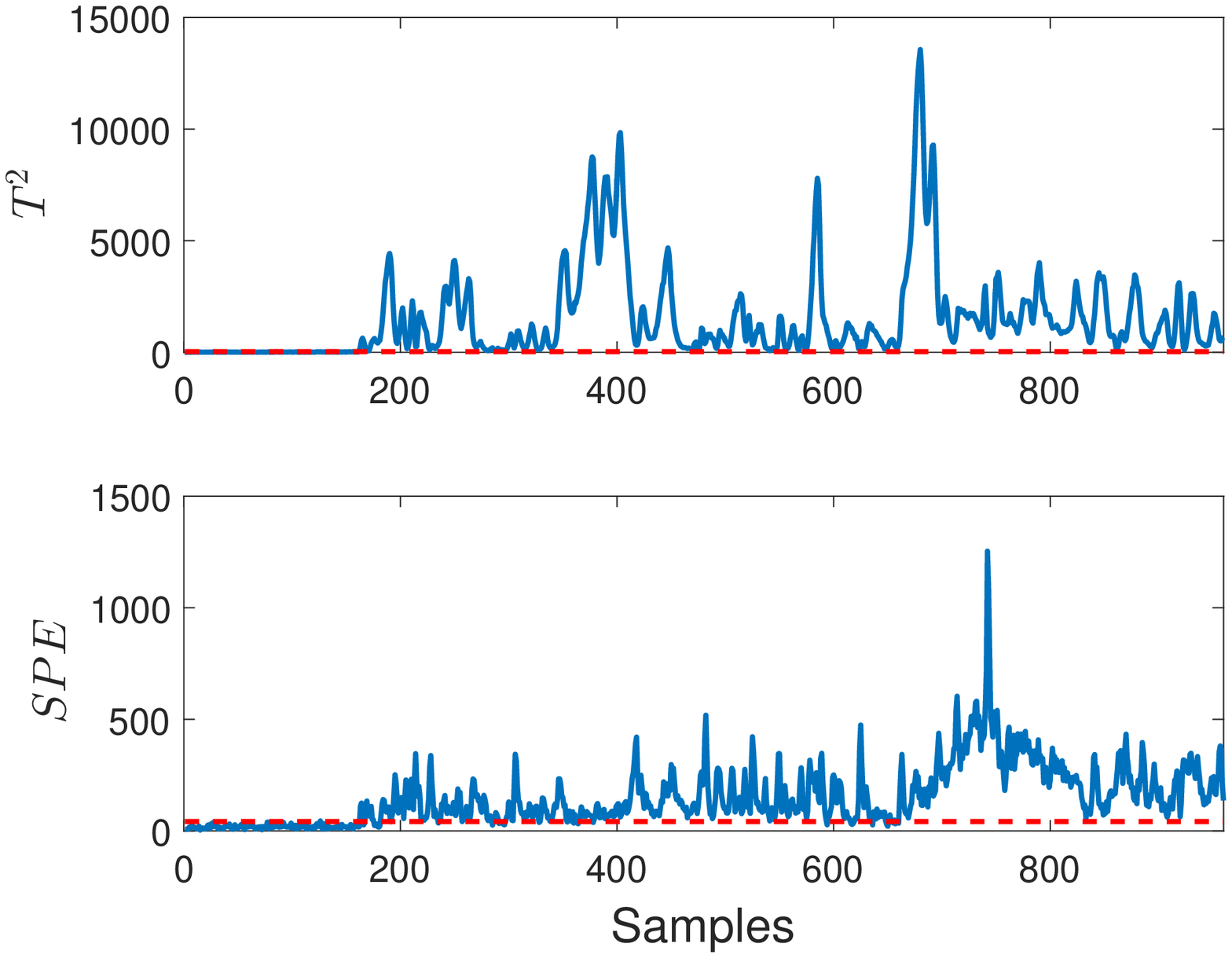}
}
\hspace{-2mm}
\subfigure[IDV14]{
\label{IDV14}
\includegraphics[width=0.315\textwidth,height=4.5cm,angle=0]{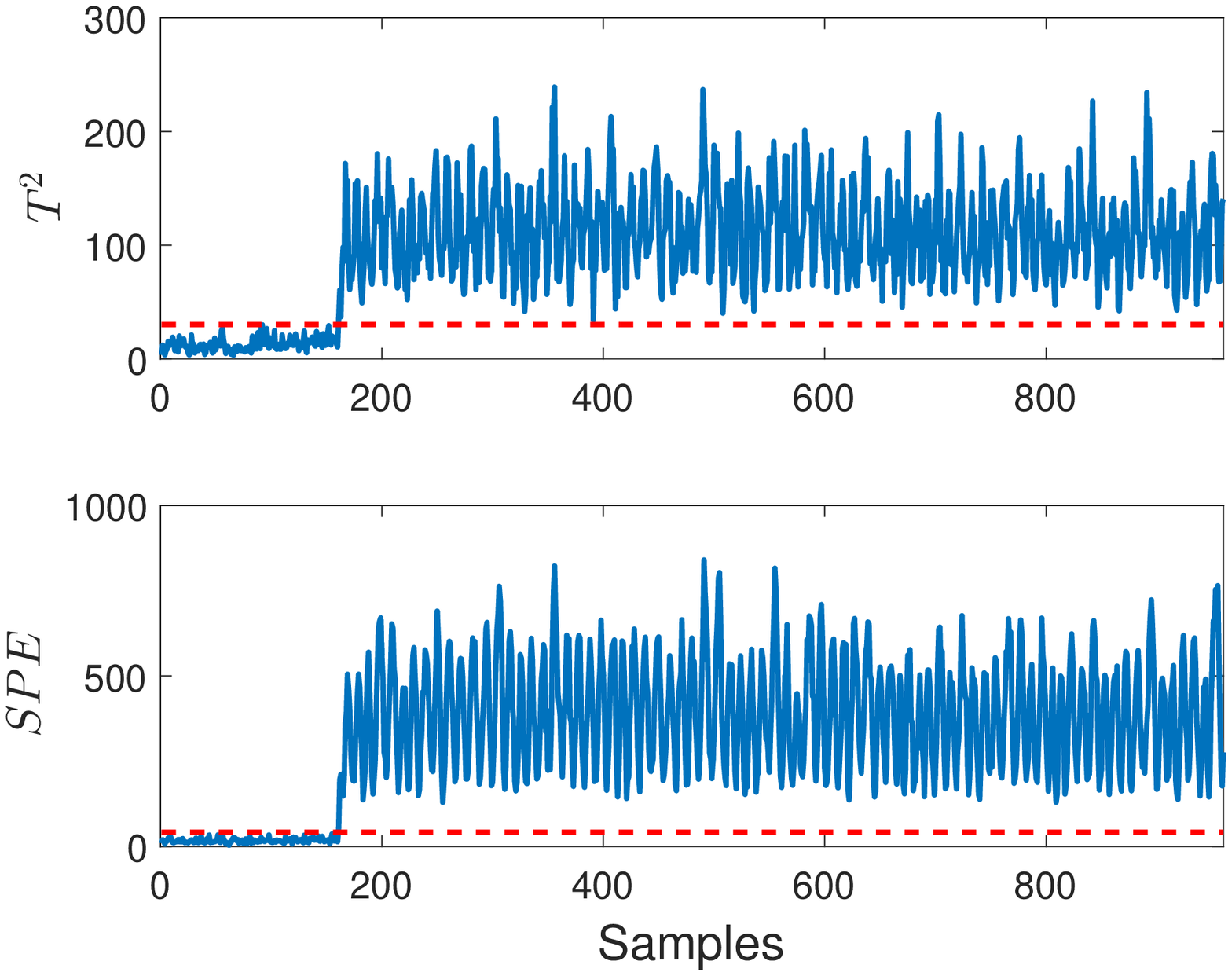}
}
\caption{Monitoring charts of typical TE faults}
\label{TE-simulation}
\end{figure*}

In this simulation case, the ID is $14$ via MLE. Regularization is selected to figure out the underlying singular problem. The monitoring consequences of three typical faults are demonstrated in Figure~\ref{TE-simulation}. It can be obviously obtained that three faults can be detected timely and accurately.  The FAR of OLPP-MLE is $0.63\%$, nearly close to $0$. The FDRs of fault IDV(1), fault IDV(12) and fault IDV(14), are $99.75\%$, $99.88\%$ and $100\%$, respectively. More specifically, two monitoring statistics can both detect these faults.

\subsection{Comparative study with other techniques}\label{comparativediscussion}
In this section, PCA, dynamic PCA (DPCA)~\cite{ku1995disturbance}, probabilistic PCA (PPCA)~\cite{tipping1999pca}, modified ICA (mICA)~\cite{constrained}, LPP,  traditional OLPP and the proposed OLPP-MLE are discussed.

 TE data is employed to demonstrate the superiority of OLPP-MLE among the approaches aforementioned. For standard LPP and OLPP, nearest neighbor dimension estimator is employed to obtain the ID and its value is 3. For mICA, the ID is 9 based on leave-one-out cross validation. Eigenvalue-based estimator is adopted for PCA, and the number of principal components (PCs) is 9. Regarding DPCA, the time lag is 2 and 28 PCs are extracted via cumulative percent variance.

\begin{table}[!tbp]
\caption{FDRs(\% ) and FARs(\%) based on TE data}\label{Table1}
\centering
{
\begin{tabular}{c c c c c c c c c}
\toprule
Fault     & PCA                & DPCA             & PPCA              & mICA           & LPP           &OLPP           & proposed \\ 
\hline
IDV(1)    & 99.88              & 99.75            & 99.75             & 98.88          & 99.75         & 99.50         & 99.75  \\  
IDV(2)    & 98.38              & 98.50            & 98.38             & 98.00          & 98.75         & 98.50         & 98.75  \\ 
IDV(3)    & 4.50               & 3.25             & 2.13              & 2.25           & 9.25          & 11.75         & 13.75  \\ 
IDV(4)    & 100                & 99.75            & 93.63             & 72.50          & 91.75         & 94.25         & 99.12  \\ 
IDV(5)    & 100                & 99.75            & 25.37             & 99.75          & 99.75         &100            & 100   \\ 
IDV(6)    & 100                & 99.75            & 100               & 100            & 100           &100            & 100    \\ 
IDV(7)    & 100                & 99.75            & 100               & 100            & 96.75         & 100           & 100   \\ 
IDV(8)    & 98.00              & 98.12            & 97.62             & 96.63          & 98.00         &98.25          & 98.25  \\
IDV(9)    & 3.50               & 3.88             & 1.87              & 3.25           & 6.50          &10.00          & 12.50  \\
IDV(10)   & 90.38              & 94.25            & 33.37             & 86.88          & 80.00         &86.88          & 91.25 \\ 
IDV(11)   & 80.63              & 93.00            & 62.50             & 56.25          & 67.87         &72.50          & 87.35  \\
IDV(12)   & 99.88              & 99.75            & 98.62             & 99.38          & 99.88         & 99.62         & 99.88  \\ 
IDV(13)   & 95.25              & 96.00            & 94.25             & 95.13          & 94.63         &95.75          & 96.13  \\ 
IDV(14)   & 100                & 99.75            & 100               & 99.88          & 100           &100            & 100   \\ 
IDV(15)   & 7.00               & 15.75            & 1.75              & 2.38           & 11.75         &12.25          & 21.13  \\  
IDV(16)   & 92.37              & 95.37            & 16.73             & 83.88          & 88.00         &84.88          & 89.50  \\     
IDV(17)   & 97.25              & 97.88            & 88.62             & 89.25          & 90.25         &89.50          & 93.63  \\  
IDV(18)   & 90.38              & 90.63            & 89.88             & 90.00          & 90.00         &90.13          & 93.37 \\ 
IDV(19)   & 94.63              & 99.62            & 20.13             & 52.25          & 74.75         &82.37          & 91.00  \\   
IDV(20)   & 91.01              & 91.01            & 41.13             & 75.50          & 81.87         &86.88          & 89.38  \\   
IDV(21)   & 57.75              & 52.25            & 40.50             & 53.63          & 45.12         &48.38          & 58.37  \\   
\hline
IDV(0)    &2.19                & 3.13             & 2.19              & 1.67           &5.37           &2.5            & 0.63  \\   
\bottomrule
\end{tabular}
}
\end{table}

\begin{table*}[htp]
\caption{A brief comparison among data-driven approaches}
\centering
\resizebox{\columnwidth}{!}{
\begin{tabular}{l l l l }
\toprule
Method   & Assumption on data         & Computational complexity                 & Parameter\\
\hline
PCA      & Gaussian distribution      & Low: 1 SVD on ${m \times m}$ matrix         & number of PCs\\
DPCA     & Same as PCA                & Medium: 1 SVD on ${hm \times hm}$ matrix    & number of PCs, $h$\\
PPCA     & Same as PCA                & High: key parameters determined by iterative EM  & number of PCs\\
mICA     & Non-Gaussian distribution  & High: cost of PCA + iterative optimization issue   & number of ICs \\
LPP      & No                         & Low: cost of PCA + adjacency graph construction     & $k$, $l$\\
OLPP     & No                         & Medium: as mentioned in Section~\ref{remarks}    &$k$, $l$\\
OLPP-MLE & No                         & Medium: as mentioned in Section~\ref{remarks}     & $k$  \\
\bottomrule
\end{tabular}
}\label{Table0}
\end{table*}

 The FARs and FDRs of these data-driven approaches are summarized in Table~\ref{Table1}. Among three manifold learning approaches, the proposed OLPP-MLE provides the optimal process monitoring performance including higher FDRs. It can be obviously discovered that the FAR of the proposed OLPP-MLE is the lowest.  OLPP-MLE algorithm dramatically outperforms others, especially for IDV(3), IDV(9), IDV(15) and IDV(21), although all of the approaches can not detect faults accurately. For other faults except IDV(19), OLPP-MLE method provides the similar or sightly better fault detection performance in comparison with the other methods.

In conclusion, from overall perspective, OLPP-MLE has the highest detection accuracy rates after the trade-off between FAR and FDRs. 

\subsection{Discussion on data-driven approaches}
This section discusses the unsupervised dimensionality reduction approaches aforementioned in several aspects, namely, basic theory, mutual relationship, data distribution, computational complexity.

PCA can extract variability information to the utmost extent, but it should follow multivariate Gaussian distribution. DPCA has the most identical procedures with PCA and time delayed vectors are considered within, which makes it considerably complicated and not appropriate for large-scale systems. PPCA was proposed based on probabilistic model, where expectation maximization (EM) is employed to estimate the principal subspace iteratively. PPCA is able to deal with missing data but it is substantially complicated~\cite{zhang2017improved}.

mICA extracts the latent statistically independent components (ICs) from non-Gaussian distribution data and can be regarded as another form of PCA~\cite{constrained}. However, the computational complexity of mICA is fairly high due to the iterative optimization issue.

Methods aforementioned are based on the global geometric properties of data. LPP can preserve local neighborhood message optimally.  However, OLPP preserves better locality performance than LPP and enables to reconstruct data conveniently. Moreover, OLPP can be regarded as another form of PCA through the specific setting\cite{he2005face}, which has been illustrated in SI file. OLPP-MLE can obtain accurate estimation of ID than traditional OLPP, thus delivering better monitoring performance, as indicated in Table~\ref{Table1}. OLPP-MLE is slightly more complicated than LPP and OLPP due to MLE. Other various virtues of the proposed method have been concluded in Section~\ref{introduction}.

A sketchy comparison among data-driven approaches aforementioned is summarized in Table~\ref{Table0}.
According to Table~\ref{Table0}, OLPP-MLE has no requirement of data distribution, and the computational complexity is moderate and acceptable. Moreover, the estimation of ID is relatively accurate and stable, thus delivering better process monitoring performance than LPP and conventional OLPP. Therefore, OLPP-MLE is a preferable choice after thorough consideration.

\section{Conclusion}\label{conclusion}

In this paper a new process monitoring approach, i.e., OLPP-MLE, has been introduced based on available sensing measurements.  The MLE is employed to estimate the intrinsic dimensionality of normal training data set, based on which OLPP is conducted to reduce dimensionality. Two test statistics are defined to monitor the model subspace and the residual subspace, and the corresponding thresholds are obtained by kernel density estimation. To deal with the singular problem in OLPP, three schemes are available to ensure the reliability of solution. Moreover, MLE provides stable and reliable estimation of intrinsic dimensionality. 
Besides, the proposed OLPP-MLE algorithm owns more discriminating capabilities for data anomaly, which may be otherwise indistinguishable via other dimensionality reduction methods. The superiority of OLPP-MLE is demonstrated by CSTR and Tennessee Eastman process in contrast with a wide range of data-driven fault detection schemes.

Actually, OLPP-MLE is based on the linear extension of Laplacian Eigenmaps and easy to reconstruct data, which shares several data representation characteristics with nonlinear techniques. In future, authors will focus on the nonlinear extension of OLPP-MLE with little artificial interference for the purpose of data-driven process monitoring. Besides, missing data and sparse data can also be considered. 

\quad

\noindent{\Large \textbf{Acknowledgment}}

\noindent This work was supported by the National Natural Science Foundation  of China (61751307, 61490701, 61473164 and 61873143) and Research Fund for the Taishan Scholar Project of Shandong Province of China.

\appendix
\section{OLPP algorithm}\label{OLPP-theory}
Algorithm of  OLPP consists of following essential steps as below~\cite{olpp}.

\textbf{Step 1: Establish the adjacency graph}

Let $G$  represent a graph with $N$ nodes. Then, the adjacency graph is established via $k$-nearest neighbor, which can be utilized to evaluate whether an edge should be put between two nodes~\cite{hasanlou2012comparative}.

\textbf{Step 2: Calculate the weights  $\boldsymbol{S}$}

${\boldsymbol S \in {\mathcal{R}^{N \times N}}}$ is a similarity matrix and obviously sparse, where $\boldsymbol S_{ij}$ measures the similarity between $\boldsymbol x_i$ and $\boldsymbol x_j$.

\textbf{Step 3: Calculate the orthogonal locality preserving projections}

The orthogonal locality preserving projections are represented by ${\left\{ {{\boldsymbol a_1}, \cdots ,{\boldsymbol a_k}} \right\}}$ and are calculated below.
\begin{equation}\label{eq2}
 {\boldsymbol A^{\left( {k- 1} \right)}} = \left[ {{\boldsymbol a_1}, \cdots ,{\boldsymbol a_k}} \right]
\end{equation}
\begin{equation}\label{eq3}
  {\boldsymbol B^{\left( {k - 1} \right)}} = {\left[ {{\boldsymbol A^{\left( {k - 1} \right)}}} \right]^{\rm T}}{\left( {{\boldsymbol X} {\boldsymbol D }{\boldsymbol X^{\rm T}}} \right)^{ - 1}}{\boldsymbol A^{\left( {k - 1} \right)}}
\end{equation}
The vectors ${\left\{ {{\boldsymbol a_1}, \cdots ,{\boldsymbol a_k}} \right\}}$ can be calculated iteratively thereinafter:

$ \bullet $ Calculate $\boldsymbol a_1$ as the eigenvector of ${{\left( {{\boldsymbol X} {\boldsymbol D} {\boldsymbol X^{\rm T}}} \right)^{ - 1}}{\boldsymbol X}{\boldsymbol L}{\boldsymbol X^{\rm T}}}$ corresponding to the smallest eigenvalue.

$ \bullet $  Calculate $\boldsymbol a_k$ as the eigenvector of the following matrix
\begin{equation}\label{eq4}
\begin{array}{l}
{\boldsymbol M^{\left(k \right)}} = \left\{ {\boldsymbol I - {{\left( {{\boldsymbol X} {\boldsymbol D} {\boldsymbol X^{\rm T}}} \right)}^{ - 1}}{\boldsymbol A^{\left( {k - 1} \right)}}{{\left[ {{\boldsymbol B^{\left( {k - 1} \right)}}} \right]}^{ - 1}}} \right.\\
\quad \quad \;\;\,\;\;\;\left. {{{\left[ {{\boldsymbol A^{\left( {k - 1} \right)}}} \right]}^{\rm T}}} \right\} \cdot {\left( {{\boldsymbol X} {\boldsymbol D} {\boldsymbol X^{\rm T}}} \right)^{ - 1}}{\boldsymbol X}{\boldsymbol L}{\boldsymbol X^{\rm T}}
\end{array}
\end{equation}
corresponding to the smallest eigenvalue of ${\boldsymbol M^{\left(k \right)}}$.

\textbf{Step 4: Orthogonal locality preserving index embedding}

Let ${\boldsymbol W_{\scriptscriptstyle {OLPI}}}  = \left[  {\boldsymbol a}_1, \cdots ,  {\boldsymbol a}_l  \right]$, then data ${\boldsymbol x}$ can be mapped as
\begin{equation}\label{eq5}
  \boldsymbol x \to \boldsymbol y = {\boldsymbol W^{\rm T}_{\scriptscriptstyle {OLPI}}} {\boldsymbol x},
\end{equation}
where $\boldsymbol y$ is an $l$-dimensional expression of raw $\boldsymbol x$.

\subsection{Relationship between OLPP and PCA\cite{he2005face}}\label{relationship}

According to ~\citeauthor{he2005face}, $\boldsymbol X \boldsymbol L \boldsymbol X^{\rm T}$ can be regarded as covariance matrix  if the Laplacian matrix $\boldsymbol L = \frac{1}{N} \boldsymbol I- \frac{1}{N^2} \boldsymbol 1 \boldsymbol 1^{\rm T}$, where $\boldsymbol I$ is the identity matrix and $\boldsymbol 1$ is a vector of all ones with proper dimension. Under the circumstances, the weight matrix $\boldsymbol S$ has simple format, i.e., $S_{ij}={1/N^2}$, $\forall i,j$. $D_{ii}=\sum\nolimits_i {{S_{ij}}}  = 1/N$.
 Let $\boldsymbol m$ denote the sample mean, i.e., $\boldsymbol m = \frac{1}{N} \sum\nolimits_i {\boldsymbol x_i}$. It can be proved as follows:
\begin{equation}
\begin{aligned}
\boldsymbol X\boldsymbol L{\boldsymbol X^{\rm T}}
&= \frac{1}{N}X\left( {\boldsymbol I - \frac{1}{N}\boldsymbol 1{\boldsymbol 1^{\rm T}}} \right){\boldsymbol X^{\rm T}}\\
&= \frac{1}{N}\sum\limits_i {{\boldsymbol x_i}\boldsymbol x_i^{\rm T}}  - \frac{1}{{{N^2}}}\left( {N \boldsymbol m} \right){\left( {N \boldsymbol m} \right)^{\rm T}}\\
&= \frac{1}{N}\sum\limits_i {\left( {{\boldsymbol x_i} -\boldsymbol m} \right){{\left( {{\boldsymbol x_i} - \boldsymbol m} \right)}^{\rm T}} + \frac{1}{N}\sum\limits_i {{\boldsymbol x_i}{\boldsymbol m^{\rm T}}} }  + \frac{1}{N}\sum\limits_i {\boldsymbol m \boldsymbol x_i^{\rm T}}  - \frac{1}{N}\sum\limits_i {\boldsymbol m{\boldsymbol m^{\rm T}}}  - \boldsymbol m{ \boldsymbol m^{\rm T}}\\
&= {\rm E}\left[ {\left( {\boldsymbol x -\boldsymbol m} \right){{\left( {\boldsymbol x -\boldsymbol m} \right)}^{\rm T}}} \right]
\end{aligned}
\end{equation}
where $ {\rm E}\left[ {\left( {\boldsymbol x -\boldsymbol m} \right){{\left( {\boldsymbol x -\boldsymbol m} \right)}^{\rm T}}} \right]$ is exactly the covariance matrix of data points.

It is evidently observed that $\boldsymbol S$ is very significant for OLPP. When global geometric structure is expected to be preserved, we just need to set $k \to \infty $ and select eigenvectors corresponding to largest eigenvalues. In this case, data points are projected along the directions of maximal variance, which implies that OLPP is equivalent to PCA in a sense. When we intend to preserve local geometric information, we need to set $k$  small enough and reserve eigenvectors associated with smallest eigenvalues. Thus, data points are projected along the directions preserving locality. The latter one is more popular and essentially 

The similarity of PCA and OLPP is that projection vectors are mutually orthogonal and can both preserve global information. However, OLPP can reserve local geometric structure, which is generally popular.


%
\end{document}